\documentclass{article}

\usepackage[top=1in,bottom=1in,left=1in,right=1in]{geometry}
\usepackage{natbib}
\usepackage[unicode=true,bookmarks=true,bookmarksnumbered=false,bookmarksopen=false,breaklinks=false,pdfborder={0 0 1},backref=false,colorlinks=true]{hyperref}
\hypersetup{citecolor=blue}
\usepackage{url}

\usepackage{amsmath}
\usepackage{mathrsfs}
\usepackage{amssymb}
\usepackage{amsthm}
\usepackage[normalem]{ulem}
\usepackage{bbm}

\newtheorem{thm}{Theorem}
\newtheorem{lem}[thm]{Lemma}

\newtheorem{prop}[thm]{Proposition}

\usepackage{float}
\usepackage{rotfloat}

\floatstyle{ruled}
\newfloat{algorithm}{tpb}{loa}
\providecommand{\algorithmname}{Algorithm}
\floatname{algorithm}{\protect\algorithmname}

\usepackage{algpseudocode,algorithm}
\algnewcommand\algorithmicinput{\textbf{Input}:}
\algnewcommand\algorithmicoutput{\textbf{Output}:}
\algnewcommand\INPUT{\item[\algorithmicinput]}
\algnewcommand\OUTPUT{\item[\algorithmicoutput]}

\usepackage{graphicx}
\usepackage[caption=false,labelformat=simple]{subfig}

\graphicspath{{/Users/yz125/Dropbox/MyFolder/Biostat-IU/Projects/Changliang/CAP_HDCov/}}

\usepackage{array}
\newcolumntype{L}[1]{>{\raggedright\let\newline\\\arraybackslash\hspace{0pt}}m{#1}}
\newcolumntype{C}[1]{>{\centering\let\newline\\\arraybackslash\hspace{0pt}}m{#1}}
\newcolumntype{R}[1]{>{\raggedleft\let\newline\\\arraybackslash\hspace{0pt}}m{#1}}

\usepackage{rotating}
\usepackage{multirow}

\usepackage{tikz}
\usetikzlibrary{shapes,arrows,backgrounds,calc,positioning,fit,petri,plotmarks}
\usetikzlibrary{arrows}

\usepackage{enumerate}
\usepackage{paralist}

\newcommand*{\affaddr}[1]{#1} % No op here. Customize it for different styles.
\newcommand*{\affmark}[1][*]{\textsuperscript{#1}}

\global\long\def\bX{\mathbf{X}}
\global\long\def\bx{\mathbf{x}}
\global\long\def\bY{\mathbf{Y}}
\global\long\def\by{\mathbf{y}}

\global\long\def\bD{\mathbf{D}}
\global\long\def\bA{\mathbf{A}}

\global\long\def\bC{\mathbf{C}}
\global\long\def\bH{\mathbf{H}}

\global\long\def\bS{\mathbf{S}}

\global\long\def\bbeta{\boldsymbol{\beta}}

\global\long\def\bSigma{\boldsymbol{\Sigma}}
\global\long\def\bgamma{\boldsymbol{\gamma}}

\global\long\def\btheta{\boldsymbol{\theta}}

\title{Covariance Regression with High-Dimensional Predictors}
\author{%
    Yuheng He\affmark[1], Changliang Zou\affmark[1], and Yi Zhao\affmark[2] \\
    \affaddr{\affmark[1]School of Statistics and Data Science, Nankai University, Tianjin, China} \\
    \affaddr{\affmark[2]Department of Biostatistics and Health Data Science, \\ Indiana University School of Medicine, USA} \\
}
\date{}

\providecommand{\keywords}[1]
{
  {\small	
  \textbf{Keywords:} #1 }
}
\begin{document}

\maketitle

\thispagestyle{empty}

\begin{abstract}
In the high-dimensional landscape, addressing the challenges of covariance regression with high-dimensional covariates has posed difficulties for conventional methodologies. This paper addresses these hurdles by presenting a novel approach for high-dimensional inference with covariance matrix outcomes. The proposed methodology is illustrated through its application in elucidating brain coactivation patterns observed in functional magnetic resonance imaging (fMRI) experiments and unraveling complex associations within anatomical connections between brain regions identified through diffusion tensor imaging (DTI). In the pursuit of dependable statistical inference, we introduce an integrative approach based on penalized estimation. This approach combines data splitting, variable selection, aggregation of low-dimensional estimators, and robust variance estimation. It enables the construction of reliable confidence intervals for covariate coefficients, supported by theoretical confidence levels under specified conditions, where asymptotic distributions are provided. Through various types of simulation studies, the proposed approach performs well for covariance regression in the presence of high-dimensional covariates. This innovative approach is applied to the Lifespan Human Connectome Project (HCP) Aging Study, which aims to uncover a typical aging trajectory and variations in the brain connectome among mature and older adults. The proposed approach effectively identifies brain networks and associated predictors of white matter integrity, aligning with established knowledge of the human brain.
\end{abstract}

%%%%%%%%%%%%%%%%%%%%%%%%%%%%%%%%%%%%%%%%%%%%%%%%%%%%%%%%%%
%========================================================%

%========================================================%

%%%%%%%%%%%%%%%%%%%%%%%%%%%%%%%%%%%%%%%%%%%%%%%%%%%%%%%%%%
%========================================================%
\clearpage
\setcounter{page}{1}

%%%%%%%%%%%%%%%%%%%%%%%%%%%%%%%%%%%%%%%%%%%%%%%%%%%%%%%%%%
% Introduction
%%%%%%%%%%%%%%%%%%%%%%%%%%%%%%%%%%%%%%%%%%%%%%%%%%%%%%%%%%
\section{Introduction}
\label{s:intro}

This manuscript investigates a high-dimensional regression problem with covariance matrices as the outcome. Assume the observed outcome data consist of $n$ subjects, each with $T_i$ observations of a $p$-dimensional random vector, denote as $\by_{i t}\in\mathbb{R}^{p}$, for $t=1, \ldots, T_i$ and $i=1, \ldots, n$. Let $\bY_i=(\by_{i1}, \ldots, \by_{i{T_i}})^\top \in \mathbb{R}^{{T_i}\times p}$ and $\bY=\{\bY_1, \ldots, \bY_n\}$. The distribution of $\by_{i t}$ is assumed to be multivariate normal with mean zero and covariance matrix $\bSigma_i$. Without loss of generality, the distribution mean is set to zero as the focus of the study is on the covariance matrices. Let $\tilde{\bx}_i \in \mathbb{R}^{q-1}$ denote the $(q-1)$ dimensional covariates of interest acquired from subject $i$. Denote $\bx_{i}=(1,\tilde{\bx}_{i}^\top)^\top\in\mathbb{R}^{q}$ the vector that includes the intercept term and $\bX=(\bx_{1},\dots,\bx_{n})^\top\in\mathbb{R}^{n\times q}$. 

The term ``high-dimensional covariance'' pertains to scenarios where $q > n$ and $q$ increases toward infinity. For covariance matrix outcomes, the following regression model introduced in \citet{zhao2021covariate} is considered. Assuming there is a linear projection, the study proposed a parsimonious model in the projection space that accounts for data heteroscedasticity via a generalized linear regression model with a logarithmic link function,
\begin{equation}
    \log \left(\boldsymbol{\gamma}^{\top} \bSigma_i \boldsymbol{\gamma}\right)=\beta_0+\tilde{\bx}_i^{\top} \boldsymbol{\beta}_1\equiv \bx_{i}^\top\bbeta,
    \label{model}
\end{equation}
where $\bgamma \in \mathbb{R}^p$ is a linear projection and $\boldsymbol{\beta}=(\beta_0, \boldsymbol{\beta}_1^\top)^\top \in \mathbb{R}^{q}$ corresponds to the model coefficients. The use of the logarithmic link ensures the positive definiteness of the covariance matrices. The primary goal is to estimate both $\bgamma$ and $\bbeta$ using the observed data, $\left\{\left(\mathbf{y}_{i 1}, \ldots, \mathbf{y}_{i T_i}\right), \mathbf{x}_i\right\}_{i=1}^n$.

This regression framework finds practical application in various domains. One notable context is its utilization for investigating the impact of covariates on brain synchronization measured by resting-state functional magnetic resonance imaging (fMRI) experiments. In the realm of fMRI studies, the utilization of covariance or correlation matrices derived from resting-state signals serves as a common approach to unveil intrinsic brain coactivation patterns. Exploring the interplay between these patterns and population or individual-level covariates emerges as a significant avenue of exploration within the field of neuroimaging research~\citep{mueller2013individual,seiler2017multivariate,zhao2021covariate}. A parallel application of this framework emerges within the realm of financial equities analysis. Consider a dataset comprising stock values, where covariance matrices calculated over specific time intervals provide insights into the level of co-movement or synchronicity among these stocks. This synchronicity is notably influenced by pertinent attributes at both the firm and market levels, including factors such as industry categorization, the firm's cash flow, stock size, and book-to-market ratio. The assessment and quantification of these intricate associations represent pivotal pursuits within the domain of financial theory~\citep{zou2017covariance}.

For a covariance regression model as in~\eqref{model}, when $q$ is less than $n$ and fixed, \citet{zhao2021covariate} conducted an initial investigation, wherein a novel approach of estimating $\bgamma$ and $\bbeta$ was introduced. This approach was grounded in likelihood principles, where iterations took place between minimizing the negative log-likelihood function within the projection space (fix $\bgamma$) and optimizing for the projection (fix $\bbeta$) with a quadratic constraint. It is worth noting that the optimization is applicable when the covariance matrix of the design matrix, $\mathbb{E}(\bx_i \bx_i^\top)$, is invertible. Consequently, when $q > n$, the above-mentioned approach becomes ill-defined, primarily attributed to the rank deficiency in the covariance matrix of the covariates. Recently, \cite{kim2024high} investigated a covariance regression model that accommodates high-dimensional covariates. They modeled $\bSigma$ as a quadratic function of covariates $\bx$, expressed as $\boldsymbol{B} \bx \bx^{\top} \boldsymbol{B}^{\top}$, where $\boldsymbol{B} \in \mathbb{R}^{p \times q}$. However, as the covariance matrix is modeled as a linear function of covariates with matrix-valued coefficients, it primarily addresses a multivariate linear regression problem, which may be less pertinent to the context under discussion here.

When encountering problems with high-dimensional covariates, one typical solution is to consider regularized estimation, including the LASSO~\citep{tibshirani1996regression}, the elastic net~\citep{zou2005regularization}, the adaptive LASSO~\citep{zou2006adaptive}, SCAD~\citep{fan2001variable}, MCP~\citep{zhang2010nearly}, ISIS~\citep{fan2008sure}, and among others. Asymptotic properties, including model selection consistency and estimation consistency, under various conditions and regularization functions, have been established \citep[see a review by][]{buhlmann2011statisticsbook}. Besides point estimation, achieving reliable inferences in high-dimensional models is also fundamental to the field. Researchers have approached this challenge from various angles. One primary direction involves making inferences based on the selected model chosen by a particular variable selection procedure. \citet{wasserman2009high} introduced a multi-stage procedure that employs data splitting to separate the steps of selection and inference. \citet{lee2016exact} developed conditional asymptotics for coefficient estimates conditional on the selected model. Another direction centers on estimating and making inferences about low-dimensional parameters within high-dimensional models. \citet{belloni2014inference} proposed a double selection procedure that involves multiple steps, rather than a single selection, to estimate and construct confidence regions for a primary regression parameter of interest. Some alternative approaches are based on penalized estimation. Notably, the bias correction method was introduced based on the LASSO proposed by~\citet{javanmard2014confidence,van2014asymptotically,zhang2014confidence}, which provides both point estimates and confidence intervals for model parameters. \citet{ning2017general} introduced hypothesis tests and confidence regions based on the decorrelated score function and test statistic. 

Each of the above-mentioned approaches has its strengths and limitations. Some methods, such as \citet{van2014asymptotically}, require precisely estimating the precision matrix of $\bx$, which is the inverse of a $q \times q$-dimensional covariance matrix, imposing a large computational burden and the assumption of sparsity in GLMs that is sometimes violated \citep{xia2020revisit}. Other methods, such as \citet{wasserman2009high} and \citet{lee2016exact}, aim for exact inference in post-selection estimates while being constrained to the correctness of model selection in the initial step. In this study, we incorporate a newly developed approach proposed by \citep{fei2021estimation} to our scenario, which utilizes sample splitting for inference with high-dimensional covariates. This approach tackles the high-dimensional estimation challenge by aggregating low-dimensional estimates, thus overcoming computational complexity issues. Importantly, the method avoids making parameter assumptions in variance estimation and does not require consistency in model selection.

The proposed covariance regression with high-dimensional covariates has two steps.
\begin{enumerate}[1.]
    \item \textbf{Penalized estimation.} It starts with estimating model parameters, $\bgamma$, and $\bbeta$, using the full dataset with regularization on the high-dimensional coefficient, $\bbeta$. 
    
    \item \textbf{High-dimensional inference.} This step employs a data splitting procedure, which divides the samples into two distinct subsets, one for dimension reduction and one for low-dimensional model fitting.
   
\end{enumerate}
The proposal is a comprehensive framework that enables estimating coefficients efficiently, conducting variable selection, and obtaining coefficient estimates for both selected and non-selected predictors. 
This approach addresses the intricacies of high-dimensional covariance regression and enhances the ability to draw meaningful inferences from the data. A pivotal assumption underlying the proposed approach is that the number of non-zero elements in the true $\bbeta$ is small, that is the sparsity assumption. 
This assumption is in line with practical implementations. For instance, the small-world topology in brain science, where human brain anatomy is economically organized~\citep{bullmore2012economy} and information processing is often highly clustered and concentrated~\citep{bassett2017small}. Another example is in the field of cancer genomics, it is plausible that a specific type of cancer is primarily associated with only a handful of oncogenes and tumor suppressor genes. 
For inference, the proposed approach is inspired by the work of \citet{fei2021estimation}, where a Splitting and Smoothing method for Generalized Linear Models (GLMs) was introduced and effectiveness under mild conditions was demonstrated. This compelling property motivates us to adapt and generalize to models with covariance outcomes, opening new avenues for statistical inference in covariance regression with high-dimensional covariates.  

The contributions of this study can be summarized into three key aspects.
\begin{itemize}
    \item \textbf{Penalized estimation and inference for covariance regression.} To our best knowledge, this is the first attempt to delve into penalized estimation and statistical inference for covariance regression under the setting of high-dimensional covariates. This pioneering work lays the foundation for addressing complex problems involving high-dimensional data and covariance matrices.
    \item \textbf{Aggregation of low-dimensional modeling.} The proposed approach tackles the challenges of high-dimensional inference by aggregating low-dimensional modeling. It significantly improves computational effectiveness compared to existing approaches, enables efficient handling of high-dimensional data, and obtains more robust estimates.
    \item \textbf{Robust variance estimation.} To ensure reliable inference, a variance estimator is developed, which utilizes the infinitesimal jackknife method following the splitting and smoothing procedure. This variance estimation approach is free from parametric assumptions, a notable advantage, and results in confidence intervals with correct coverage probabilities.
\end{itemize}

The rest of the paper is organized as follows. Section~\ref{sec: estimation} introduces the proposed penalized estimator of model coefficients and imposed assumptions for estimation and inference and studies the asymptotic properties. In Section~\ref{sec: inference}, a Splitting and Smooth method for statistical inference is introduced and asymptotic properties are derived. 
Section~\ref{sec: simulation} demonstrates the good performance of the proposed approach via simulation studies. 
Section~\ref{sec: application} articulates an application to the Lifespan Human Connectome Project (HCP) Aging Study. 
Section~\ref{sec: discussion} summarizes the manuscript with discussions.
The supporting information also collects the technical proof of the theorems in the main text, additional theoretical results, and additional data analysis results.
%%%%%%%%%%%%%%%%%%%%%%%%%%%%%%%%%%

%%%%%%%%%%%%%%%%%%%%%%%%%%%%%%%%%%
\section{Estimation}
\label{sec: estimation}

%==========================================
\subsection{Method}
\label{sub:est_method}

Given the regression model~\eqref{model}, it is proposed that an estimation of the parameters can be obtained by solving the following optimization problem. Specifically, the optimization problem is formulated to minimize the regularized negative likelihood function: 

\begin{eqnarray}
    \underset{\boldsymbol{\beta}, \boldsymbol{\gamma}}{\operatorname{minimize}} && \ell(\boldsymbol{\beta}, \boldsymbol{\gamma}):=\frac{1}{2} \sum_{i=1}^{n} T_{i}\left\{\left(\mathbf{x}_i^{\top} \boldsymbol{\beta}\right)+ \boldsymbol{\gamma}^{\top} \hat{\bSigma}_i \boldsymbol{\gamma} \cdot \exp \left(-\mathbf{x}_i^{\top} \boldsymbol{\beta}\right)\right\} + \lambda\| \bbeta\|_1, \nonumber \\
    \text{such that} && \boldsymbol{\gamma}^{\top} \mathbf{H} \boldsymbol{\gamma}=1. \label{obj}
\end{eqnarray}
One may express the objective function as $\ell(\boldsymbol{\beta}, \boldsymbol{\gamma})=\mathcal{L}_n(\boldsymbol{\beta}, \boldsymbol{\gamma})+\mathcal{P}(\boldsymbol{\beta})$, wherein the first component denotes the negative log-likelihood and the second denotes the regularization.

The first component employs an estimator, denoted by $\hat{\bSigma}_i$, to approximate the true covariance matrix, $\bSigma_i$. Under the normally assumption and likelihood criterion, $\hat{\bSigma}_i=\mathbf{S}_i=\sum_{t=1}^{T_i} \mathbf{y}_{i t} \mathbf{y}_{i t}^{\top} / T_i$ is the sample covariance matrix as introduced in \citet{zhao2021covariate}. 
To identify $\bgamma$, a quadratic constraint concerning a positive definite matrix ($\bH$) is necessary, analogous to the principal component analysis (PCA) principle. With a likelihood-based objective function,  $\bH$ is set to be the average of the $\hat{\bSigma}_{i}$'s with $\mathbf{H}=\sum_{i=1}^n T_i \hat{\bSigma}_i / \sum_{i=1}^n T_i$ to incorporate distributional information~\cite[see also the discussion in][]{zhao2021covariate}.

When the value of $q$ significantly exceeds that of $n$ (a high-dimensional scenario), the problem becomes challenging due to the rank deficiency of the design matrix. To address this issue, a regularization component, $\mathcal{P}(\bbeta)$, is introduced. Examples include the $\ell_1$-regularization~\citep{tibshirani1996regression} and its extensions. Existing findings also guarantee both estimation consistency and selection consistency under mild conditions. A generalization to the current model setting with covariance matrices as the outcome will be comprehensively discussed in Section~\ref{sub:est_asmp}.

The present study considers an optimization problem of~\eqref{obj}, which exhibits biconvexity. Following the approach utilized in \citet{zhao2021covariate}, we consider a high-dimensional block coordinate descent algorithm that is well-suited for fitting model parameters, $\boldsymbol{\gamma}$ and $\boldsymbol{\beta}$. The procedure is described in Algorithm \ref{alg1}. To enhance the likelihood of escaping from local minima, we recommend adopting a randomized approach to selecting initial values, followed by selecting the estimate that yields the lowest value of the objective function.

Similar to Algorithm 2 in \citet{zhao2021covariate}, when there are higher-order components, it is possible to first remove the identified components and utilize the updated data for estimating the succeeding component. This estimation involves incorporating an orthogonality constraint, ensuring the newfound component is orthogonal to the previously identified ones. To determine the number of projections, the same metric introduced by \citet{zhao2021covariate} is utilized, which quantifies the ``deviation from diagonality'' of the projected matrices. Let $\Gamma^{(k)} \in \mathbb{R}^{p \times k}$ denote the first $k$ estimated components, the average deviation from diagonality is defined as
$$
\operatorname{DfD}\left(\Gamma^{(k)}\right)=\prod_{i=1}^n\left(\frac{\operatorname{det}\left\{\operatorname{diag}\left(\Gamma^{(k) \top} \mathbf{S}_i \Gamma^{(k)}\right)\right\}}{\operatorname{det}\left(\Gamma^{(k) \top} \mathbf{S}_i \Gamma^{(k)}\right)}\right)^{T_i / \sum_i T_i},
$$
where $\operatorname{diag}(\mathbf{A})$ represents a diagonal matrix comprising the diagonal elements of a square matrix $\mathbf{A}$, and $\operatorname{det}(\mathbf{A})$ is the determinant of $\mathbf{A}$. If $\Gamma^{(k)}$ is a shared diagonalization of $\mathbf{S}_i$'s, meaning $\Gamma^{(k) \top} \mathbf{S}_i \Gamma^{(k)}$ is a diagonal matrix, for $\forall~ i=1, \ldots, n$, then $\operatorname{DfD}\left(\Gamma^{(k)}\right)=1$.  In practice, the selection of $k$ can occur before DfD significantly deviates from one or before a sudden increase takes place.

\begin{algorithm}
	%\textsl{}\setstretch{1.8}
	\caption{The optimization algorithm for problem~\eqref{obj}.}
	\label{alg1}
	\begin{algorithmic}[1]
		\INPUT 
                   $\mathbf{Y}$: a list of data where the $i$th element a $T_i \times p$ data matrix;
                   
                   $\mathbf{X}$: an $n \times q$ matrix of covariates with the first column of ones.
		
            \State Initialization: $\boldsymbol{\beta}^{(0)}, \boldsymbol{\gamma}^{(0)}$
            \Repeat  \quad for iteration $s=0,1,2,\ldots$, 
            given $\left(\boldsymbol{\beta}^{(s)}, \boldsymbol{\gamma}^{(s)}\right)$, for the $(s+1)$th step:
            
		\State(1) update $\bbeta$ by solving 
            \begin{equation*}
		    \frac{1}{2} \sum_{i=1}^n T_{i}\left\{\left(\mathbf{x}_i^{\top} \boldsymbol{\beta}\right)+ {\boldsymbol{\gamma}^{(s)}}^{\top} \mathbf{S}_i \boldsymbol{\gamma}^{(s)} \cdot \exp \left(-\mathbf{x}_i^{\top} \boldsymbol{\beta}\right)\right\} + \lambda\| \bbeta\|_1;
		\end{equation*}
		\State(2) update $\bgamma$ by solving 
                   \begin{eqnarray*}
                    \underset{\bgamma}{\operatorname{minimize}} && \bgamma^\top\left\{\frac{1}{2} \sum_{i=1}^n T_{i} \exp \left(-\mathbf{x}_i^{\top} \boldsymbol{\beta}^{(s+1)}\right)\bS_{i}\right\} \bgamma, \\
                   \text {such that} && \boldsymbol{\gamma}^{\top} \mathbf{H} \boldsymbol{\gamma}=1;
                   \end{eqnarray*}
		\Until  the objective function in~\eqref{obj} converges.
            \State Consider a random series of initializations, repeat Steps 1--5, and choose the results with the minimum objective value.
            \OUTPUT $\hat{\boldsymbol{\beta}}, \hat{\bgamma}$
	\end{algorithmic}  
\end{algorithm}

%==========================================

%==========================================

\subsection{Asymptotic properties}
\label{sub:est_asmp}

In this section, we establish the estimation consistency and model selection consistency of the proposed approach. We denote the true model parameter as $\bbeta^*$ and the model space under regularization $\mathcal{P}$ (for example, the $\ell_1$-norm) as $\mathcal{M}=\left\{\bbeta \in\mathbb{R}^p \mid( \bbeta)_{S^c}=\boldsymbol{\mathrm{0}}\right\}$, where $S$ is the support of $ \bbeta^*$ and $S^c$ is the complement of $S$. To ensure the desired theoretical properties, we impose the following regularity conditions.
\begin{itemize}
    \item[(A1)] (Bounded observations) For a random feature vector $\tilde{\bx} \in \mathbb{R}^{q-1}, \|\tilde{\bx}\|_{\infty} \leq C_0 \text { and } \mathbb{E}(|\bY|)<\infty$.
    \item[(A2)] (Bounded eigenvalues) The eigenvalues of $\Omega=\mathbb{E}\left(\mathbf{x}_{i} \mathbf{x}_{i}^\top\right)$, denoted as $\mu(\Omega)$ , are bounded below and above by constants $c_{\min }, c_{\max }$, such that
$$
0<c_{\min } \leq \mu_{\min }(\Omega)<\mu_{\max }(\Omega) \leq c_{\max }<\infty.
$$
    \item[(A3)] (Irrepresentability) For $\tau \in(0,1)$,

$$
\left\| \mathbf{X}_{S^c}^{\top}\left( \mathbf{X}_S^{\top}\right)^{\dagger} \operatorname{sign}\left( \boldsymbol{\beta}^*_S\right)\right\|_{\infty} \leq 1-\tau,
$$
\end{itemize}
where $\bX=(\bx_{1},\dots,\bx_{n})^\top\in\mathbb{R}^{n\times q}$, $\bX_{S^{c}}$ is the columns of $\bX$ in $S^{c}$, $\bX_{S}$ is the columns of $\bX$ in $S$, and $\bbeta_{S}^{*}$ is the corresponding elements of $\bbeta^{*}$ in $S$. For a matrix $\bA \in \mathbb{R}^{p \times p}$, $\bA^{\dagger}$ is \textit{Moore-Penrose pseudoinverse} and sign($\cdot$) is the sign function.

\begin{itemize}
    \item[(A4)] (Common eigenstructure) All the covariance matrices share the same set of eigenvectors, denoted as $\Gamma_{i}$ and $\Gamma_i=\Gamma$, for $i = 1, \ldots, n$. For each $\bSigma_i$, there exists (at least) a column, indexed by $j_i$, such that $\bgamma = \boldsymbol{\gamma}_{i {j_i}}$ and Model \eqref{model} is satisfied.
\end{itemize}

The random design assumption postulated in Assumption (A1) is a fundamental and prevalent condition in the literature~\citep{song2020convex}. Additionally, it is reasonable to restrict the range of predictors via appropriate data preprocessing techniques. Assumption (A2) regulates the covariance matrix of $\mathbf{x}$, denoted as $\Omega$. It also serves as the expected Hessian matrix of the considered objective function, which is likelihood-based. Assumption (A2) is a standard regularity condition, requiring that the eigenvalues of the expected Hessian matrix remain bounded away from zero and infinity. This condition is necessary to guarantee the existence of the maximum likelihood estimate and to ensure the stability of the optimization algorithm. Assumption (A3) pertains to the irrepresentability condition which necessitates that active predictors are not excessively aligned with inactive predictors. Ideally, the inactive predictors are orthogonal to the active predictors. However, the high dimensionality of the data makes this an impractical assumption. Therefore, Assumption (A3) relaxes the orthogonality requirement to near-orthogonality, enabling greater flexibility in model specification. Assumption (A4) posits that all covariance matrices share an identical eigenspace, irrespective of the order. For $i=1, \ldots, n$, it is assumed that $\bSigma_i$ has the eigendecomposition of $\bSigma_i=\Gamma_i \Lambda_i \Gamma_i^{\top}$, where $\Lambda_i=\operatorname{diag}\left\{\lambda_{i 1}, \ldots, \lambda_{i p}\right\}$ is a diagonal matrix and $\Gamma_i=$ $\left(\boldsymbol{\gamma}_{i 1}, \ldots, \boldsymbol{\gamma}_{i p}\right)$ is an orthonormal rotation matrix; $\left\{\lambda_{i 1}, \ldots, \lambda_{i p}\right\}$ are the eigenvalues and the columns of $\Gamma_i$ are the corresponding eigenvectors. Under Assumption (A4), we investigate the theoretical outcomes of the proposed method under the assumption of common principal components.

To begin with, we aim to establish a connection between Assumptions (A1), (A2), and the well-known condition of Restricted Strong Convexity (RSC) introduced by \citet{negahban2012unified}. The basic concept underlying the definition of RSC is the relaxation of strong convexity, which is unable to hold under the high-dimensional setting due to rank deficiency, resulting in vanishing curvature of $\mathcal{L}_{n}$ along certain directions. The RSC condition ensures the availability of gradient information for the algorithm to navigate towards the optimal solution $\bbeta^*$ despite the absence of strong convexity.

\begin{prop}\label{prop:beta_RSC}
Assume Assumptions (A1) and (A2) are satisfied and $q \gg n$ and $\log{q}/M_n =o(1)$, where $M_n=\sum_{i=1}^n T_i$. Then for any given $\epsilon>0$, there exist positive constants $\alpha_1$, $\alpha_2$, and $C$ such that for $n \geq C \log (1 / \epsilon)$, it holds with probability at least $1-\epsilon$ that
\begin{equation}\label{eq:RSC}
\left(\nabla \mathcal{L}_n(\bbeta)-\nabla \mathcal{L}_n\left(\bbeta^*\right)\right)^{\top}\left(\bbeta-\bbeta^*\right) \geq \alpha_1\left\|\bbeta-\bbeta^*\right\|_2^2-\alpha_2 \sqrt{\frac{\log q}{M_n}}\left\|\bbeta-\bbeta^*\right\|_1^2, \quad \forall~\bbeta \in \mathbb{B}_2\left(r ; \bbeta^*\right),
\end{equation}
where $\mathbb{B}_{2}(r;\bbeta^{*})$ is a compact region containing $\bbeta^{*}$ with radius $r$. %\change{[Need to demonstrate what $\mathbb{B}_{2}(1;\bbeta^{*})$ means.]}
Inequality~\eqref{eq:RSC} is the restricted strong convexity. 
\end{prop}

Then We first discuss the asymptotic property of our $\bbeta$ estimator given the true $\bgamma$. Incorporating the RSC and Assumption (A3), the present study advances a novel theorem that closely resembles Theorem 3.4 in~\citet{lee2015model}.

\begin{thm}\label{thm:consistent}
Under RSC and Assumption (A3), for some $\tau$ defined in Assumption (A3), constants $\kappa, c_1, c_2$ and $\lambda=(3\kappa c_1/\tau)\sqrt{\log{q} / c_2 M_n}$ with $M_n=\sum_{i=1}^n T_i$, the estimator of $\bbeta$ under $\mathcal{P}$ is unique, and with probability at least $1-2 q^{-5/4}$,
    \begin{itemize}
        \item[(1)] estimation consistent: $$
        \left\|\hat{\bbeta}-\bbeta^*\right\|_2 \leq \frac{6}{\alpha_1}\frac{\kappa}{\tau}c_1\left(\sqrt{|S|}+\frac{\tau}{2\kappa}\sqrt{|S|}\right)\sqrt{\frac{\log q}{c_2 M_n}},$$
        \item[(2)] model selection consistent: $$
        \hat{\bbeta} \in \mathcal{M},$$
    \end{itemize}
    where $|S|$ is the cardinality of set $S$.
\end{thm}
The proof of Theorem \ref{thm:consistent} and values of $\kappa$ are provided in Supporting Information Section~\ref{appendix:sub:proof_consistent}.

\subsection{Extension}
\label{sub:Ext}

In practice, based on domain knowledge, structural constraints on the predictors are sometimes enforced, instead of imposing pure sparsity. Assuming the predictor structure can be captured by a matrix $\bD\in\mathbb{R}^{r\times q}$, where $r$ is the number of structural constraints, the imposed regularization is then $\mathcal{P}(\bbeta)=\| \bD \bbeta\|_1$. We refer to these problems as the generalized lasso~\citep{tibshirani2011solution}. Example choices of $\bD$ include the fused lasso, trend filtering, wavelet smoothing, and a method for outlier detection. 
In our neuroimaging application, the goal is to evaluate whether brain white matter integrity regulates brain functional connectivity. Local constancy and smoothness within each functional module are imposed on the structural imaging data~\citep{grosenick2013interpretable}, and the fused lasso regularization is particularly well-suited.

We consider the following optimization problem under structured constraints. 
\begin{eqnarray*}
    \underset{\boldsymbol{\bbeta}, \boldsymbol{\bgamma}}{\operatorname{minimize}} && \ell(\boldsymbol{\beta}, \boldsymbol{\gamma})=\frac{1}{2} \sum_{i=1}^n T_{i}\left\{\left(\mathbf{x}_i^{\top} \boldsymbol{\beta}\right)+ \boldsymbol{\gamma}^{\top} \mathbf{S}_i \boldsymbol{\gamma} \cdot \exp \left(-\mathbf{x}_i^{\top} \boldsymbol{\beta}\right)\right\} + \lambda\|\bD \bbeta\|_1, \\
    \text{such that} && \boldsymbol{\gamma}^{\top} \mathbf{H} \boldsymbol{\gamma}=1. 
\end{eqnarray*}
When the matrix $\bD$ is invertible, let $\btheta\equiv\bD\bbeta$ ($\bbeta=\bD^{-1}\btheta$), it reparameterizes the problem with $\btheta$ and the same approaches can be utilized followed by the asymptotic consistency. When $\bD$ is not invertible, to guarantee the theoretical properties of the generalized lasso, an adapted version of the third condition (A3) is introduced.
% , together with the first two assumptions (A1) and (A2) remaining the same, 

Let $\bbeta^*$ denote the true model parameter and $\mathcal{M^{\prime}}=\left\{\bbeta \in\mathbb{R}^q \mid(\mathbf{D} \bbeta)_{{S^{\prime}}^c}=0\right\}$ denote the model space under regularization $\mathcal{P}(\bbeta)=\|\bD\bbeta\|_{1}$, where $S^{\prime}$ is the support of $\mathbf{D} \bbeta^*$ and ${S^{\prime}}^c$ is the complement of $S^{\prime}$. The following irrepresentability condition is assumed.

\begin{itemize}
    \item[(A$3^\prime$)] (Irrepresentability) For $\tau \in(0,1)$,
$$
\left\|\mathbf{D}_{{S^{\prime}}^c} \mathbf{X}^{\top}\left(\mathbf{D}_{S^{\prime}} \mathbf{X}^{\top}\right)^{\dagger} \operatorname{sign}\left\{\left(\mathbf{D} \boldsymbol{\beta}^*\right)_{S^{\prime}}\right\}\right\|_{\infty} \leq 1-\tau.
$$
\end{itemize}

This adapted version of the irrepresentability condition requires that the active predictors, concerning $\bD$, should not be overly well-aligned with the inactive predictors. The following theorem is an analogy to Theorem~\ref{thm:consistent} for the generalized lasso.

\begin{thm}\label{thm:consistent_general}
    Under RSC and Assumption (A$3^\prime$),  for some $\tau$ defined in Assumption (A$3^{\prime}$), constants $\kappa^{\prime}, c_1^{\prime}, c_2^{\prime}$ and $\lambda=(3\kappa^{\prime} c_1^{\prime}/\tau)\sqrt{\log{q} / c_2^{\prime} M_n}$ with $M_n=\sum_{i=1}^n T_i$, the estimator of $\bbeta$ under $\mathcal{P}$ is unique, and with probability at least $1-2 q^{-5/4}$,
    \begin{itemize}
        \item[(1)] estimation consistent:
$$
\left\|\hat{\bbeta}-\bbeta^*\right\|_2 \leq \frac{6}{\alpha_1}\frac{\kappa^{\prime}}{\tau}c_1^{\prime}\left(\sqrt{|S^{\prime}|}+\frac{\tau}{2\kappa^{\prime}}\sqrt{|S^{\prime}|}\right)\sqrt{\frac{\log q}{c_2^{\prime} M_n}},
$$
        \item[(2)]  model selection consistent:
$$
\hat{\bbeta} \in \mathcal{M^{\prime}}.
$$
    \end{itemize}
\end{thm}

%==========================================

%%%%%%%%%%%%%%%%%%%%%%%%%%%%%%%%%%%%%%%%%%%%%%%%%%%%%%%%%%

\section{Inference}
\label{sec: inference}

%==========================================
\subsection{Method}
\label{sub:inf_method}
In general, it is often necessary to make statistical inferences on the estimated model coefficients, such as to evaluate the significance of the effect of a covariate or to obtain confidence intervals of the coefficient estimates. However, when faced with high-dimensional covariates, the resulting estimates are typically biased, rendering additional challenges. This is evident in the use of a lasso-type of estimation in the preceding section. To overcome this obstacle, we utilize a ``split-and-smooth'' approach~\citep{fei2021estimation}, which combines low-dimensional estimations with prior variable selection to correct for bias in the high-dimensional problem. In the following discussion of inference on $\bbeta$, it is assumed that the projection vector, $\bgamma$, is given, either prespecified or using the estimate from the full dataset. Based on the formulation in Model~\eqref{model}, define $z_{i}\equiv \sum_{t=1}^{T_{i}}(\bgamma^\top\by_{it})^{2}$, and consider it as the response variable. Notably, after standardization, $z_i$ has been shown to follow a chi-squared distribution, a member of the exponential distribution family. Consequently, fitting Model~\eqref{model} between the predictor $\bx_i$ and $z_i$ corresponds to solving a generalized linear model (GLM) problem. 

We consider $n$ subjects of observed data $\mathbf{C}^{(n)}=(\mathbf{Y},\mathbf{X})=\{(\bY_i,\mathbf{x}_i)\mid 1,\ldots, n\}$, where $\bY=\{\bY_1, \ldots, \bY_n\}$, $\bY_i=(\by_{i1}, \ldots, \by_{i{T_i}})^\top \in \mathbb{R}^{{T_i}\times p}$ and $\bX=(\bx_{1},\dots,\bx_{n})^\top\in\mathbb{R}^{n\times q}$. For Model~\eqref{model}, a sparsity condition is imposed where the number of nonzero coefficients is small relative to the sample size. More details will be explained in the upcoming subsection. Considering a high-dimensional problem, a data-splitting approach is employed as an initial step. For instance, the observed dataset is divided into two mutually exclusive subsets, denoted as $\mathbf{C}_1=(\bY^{1},\bX^{1})$ and $\mathbf{C}_2=(\bY^{2},\bX^{2})$, of sizes $n_1$ and $n_2$ respectively, with $n_1+n_2=n$. The former subset, $\mathbf{C}_1$, is used for dimension reduction, where a variable selection technique (denoted as $\mathcal{S}_\lambda$, where $\lambda$ is the tuning parameter used in the technique) is implemented and a subset of predictors, denoted as $\hat{S}\subseteq\{1,\dots,q\}$, is selected. The second dataset, $\mathbf{C}_2$, is utilized to fit low-dimensional models of~\eqref{model} using $\mathbf{Y}^2$ and $\mathbf{X}_{\hat{S}_{+j}}^2$, where $\hat{S}_{+j}=\{j\}\cup \hat{S}$, for $j=1,2,\ldots,q$. Let $\tilde{\bbeta}_{\hat{S}_{+j}}$ denote the corresponding maximum likelihood estimate and $\tilde{\beta}_j=\left(\tilde{\bbeta}_{\hat{S}_{+j}}\right)_j$ denote the element corresponding to the $j$th predictor. Note that, the first column of $\bX$ are ones, thus, $\tilde{\beta}_{1}$ is the estimate of the intercept. The estimate of the model coefficients from one data split, namely single-splitting estimator, is then obtained and denoted as $\tilde{\bbeta}=(\tilde{\beta}_{1},\dots,\tilde{\beta}_{q})^\top\in\mathbb{R}^{q}$.

Utilizing a single-split estimator is recognized with high variability, primarily due to the nature of random data splitting. This heightened variability poses a challenge in distinguishing genuine signals from noise, a recurrent issue encountered in bagging algorithms \citep{buhlmann2002analyzing}. To address this issue, we are considering implementing a multi-splitting strategy. 
We repeat the above single-splitting procedure for $B$ times, where $B$ is a large number. Denote $\bC_{1}^{b}$ and $\bC_{2}^{b}$ as the two subsets of the $b$th splitting with $|\bC_{1}^{b}|=n_{1}$ and $|\bC_{2}^{b}|=n_{2}$, for $b=1,\dots,B$. We apply $\mathcal{S}_{\lambda}$ on $\bC_{1}^{b}$ to obtain the estimate of the active set, $\hat{S}^{b}$, and yield the estimate of model coefficients by fitting a low-dimensional model using $\bC_{2}^{b}$, denoted as $\tilde{\bbeta}^{b}=(\tilde{\beta}_{1}^{b},\dots,\tilde{\beta}_{q}^{b})^\top$. Then, these $B$ estimates of model coefficients are averaged over all splits to attain a smoothed coefficient estimate, defined as
\begin{equation}
    \hat{\bbeta}=\left(\hat{\beta}_{1}, \dots, \hat{\beta}_{q}\right)^\top, \text { where } \hat{\beta}_j=\frac{1}{B} \sum_{b=1}^B \tilde{\beta}_j^b , \text{ for } j=1,\dots,q.
    \label{smooth}
\end{equation}
The process for the multi-splitting estimator $\hat{\beta}_j$ is detailed in Algorithm~\ref{alg2}. The subsequent subsection will delve into the theoretical properties of this procedure, enabling the construction of a confidence interval for $\hat{\beta}_j$ to facilitate inference.

\begin{algorithm}
	%\textsl{}\setstretch{1.8}
	\caption{The Multi-splitting Estimator}
	\label{alg2}
	\begin{algorithmic}[1]
		\INPUT 
                   $\mathbf{Y}$: a list of data, where the $i$th element is a $T_i \times p$ matrix;
                   
                   $\mathbf{X}$: an $n \times q$ matrix of covariates with the first column of ones; 
                   
                   $\hat{\bgamma}$: a $p$-dimensional vector denoting the prespecified or estimated projection vector obtained from Algorithm~\ref{alg1};
                   
                   $\mathcal{S}_\lambda$: a variable selection procedure;
                   
                   $n_1, n_2$: sample sizes for splitting;
                   
                   $B$: the number of random splits.
		
            \For{ $b=1,2, \ldots, B$}
            
                \State Split the samples into $\mathbf{C}_1^{b}$ and $\mathbf{C}_2^{b}$, with $\left|\mathbf{C}_1^{b}\right|=n_1$ and $\left|\mathbf{C}_2^{b}\right|=n_2$.
                \State With given $\hat{\bgamma}$, apply $\mathcal{S}_\lambda$ to $\mathbf{C}_1^{b}$ to select predictors indexed by $\hat{S}^{b} \subset\{1, \ldots, q\}$.
            
		          \For{$j=1, \ldots, q$}
            
                      \State With $\hat{S}_{+j}^{b}=\{j\} \cup \hat{S}^{b}$, fit model~\eqref{model} using $\bY^{b2}$ and $\bX_{\hat{S}_{+j}}^{b2}$ with $\bgamma=\hat{\bgamma}$ and obtain the estimate of model coefficients, denoted as $\tilde{\bbeta}_{\hat{S}_{+j}}^{b}$.
                      \State Let $\tilde{\beta}_j^b=\left(\tilde{\boldsymbol{\beta}}_{\hat{S}_{+j}^b}^{b}\right)_j$, which is the coefficient for predictor $\mathbf{X}_j$.
                    \EndFor
                \State Set $\tilde{\boldsymbol{\beta}}^b=\left(\tilde{\beta}_1^b, \dots, \tilde{\beta}_q^b\right)^\top$.
            \EndFor
            \State Compute $\hat{\boldsymbol{\beta}}=\left(\hat{\beta}_1, \dots, \hat{\beta}_q\right)^\top$, where $\hat{\beta}_j=\frac{1}{B} \sum_{b=1}^B \tilde{\beta}_j^b$
            \OUTPUT $\hat{\boldsymbol{\beta}}$.
	\end{algorithmic}  
\end{algorithm}

%==========================================
%==========================================
\subsection{Asymptotic properties}
\label{sub:inf_asmp}

According to the model selection consistency outlined in Theorem~\ref{thm:consistent}, the sure screening property is readily satisfied under regularity conditions. Assuming $\gamma$ is known, similar properties as those in \citet{fei2021estimation} can be derived. Define the \textit{observed information} as $\hat{I}(\bbeta)=1/2M_n \sum_{i=1}^n T_i \exp \left(-\mathbf{x}_i^{\top} \boldsymbol{\beta}\right)\left(\boldsymbol{\gamma}^{\top} \mathbf{y}_{i t}\cdot \right. $ $\left. \mathbf{y}_{i t}^{\top} \boldsymbol{\gamma}\right) \mathbf{x}_i \mathbf{x}_i^{\top}$, the \textit{expected information} as $I^*=\mathbb{E}\{\hat{I}(\bbeta^*)\}$, where $\bbeta^{*}$ is the true model parameter. For $S \subseteq \{1,2,\ldots, q\}$, the support of $\bbeta^{*}$, we have an estimation $\hat{S} \supseteq S$ and $\bx_{iS}, \bbeta_S$ are subvectors of $\bx_i$ and $\bbeta$ only containing elements indexed by $S$, respectively. Moreover, $S_{+j}=S \cup \{j\}$ and $S_{-j}=S \backslash \{j\}$. Similarly, define the \textit{observed sub-information} by $\hat{I}_{\hat{S}} (\bbeta_{\hat{S}} )=1/2M_n \sum_{i=1}^n T_i \exp \left(-\mathbf{x}_{i\hat{S}}^{\top} \boldsymbol{\beta}_{\hat{S}}\right)\left(\boldsymbol{\gamma}^{\top} \mathbf{y}_{i t}\cdot \right.$$\left. \mathbf{y}_{i t}^{\top} \boldsymbol{\gamma}\right) \mathbf{x}_{i\hat{S}} \mathbf{x}_{i\hat{S}}^{\top}$ and the \textit{expected sub-information} by $I_{\hat{S}} = \mathbb{E} \{\hat{I}_{\hat{S}}(\bbeta^*_{\hat{S}} )\}=I^*_{\hat{S}}$, which is the submatrix of $I^*$ with rows and columns indexed by $\hat{S}$.

The asymptotic properties of the single-splitting estimator are outlined in Theorem~\ref{thm:asy_single}, demonstrating its consistency and normality. However, due to its utilization of only $n_1$ subjects, it exhibits reduced efficiency. In contrast, a multi-splitting strategy is employed to address the high variability associated with data splitting, yielding a smoothed estimator denoted as $\hat{\bbeta}$. The subsequent Theorem~\ref{thm:asy_multi} details the asymptotic properties of this smoothed estimator.
\begin{thm}\label{thm:asy_single}
When the dimension of the predictors is divergent with sample size $n$, consider the single-splitting estimator $\tilde{\bbeta}=(\tilde{\beta_1}, \dots, \tilde{\beta_q})^\top$. With known $\boldsymbol{\gamma}$, denote $m=|\hat{S}|$ and $\tilde{\sigma}_j^2=\left(\left\{I^*_{\hat{S}_{+j}}\right\}^{-1}\right)_{jj}$, for $j \in \{1,\ldots,q\}$. Let $T=\min_i T_i$ and $M_n^1=\sum_{i=1}^n T_i$. As $n, T \rightarrow \infty$, 
\begin{enumerate}[(1)]
    \item 
        \begin{equation*}
\left\|\tilde{\bbeta}_{S_{+j}}-\bbeta_{S_{+j}}^*\right\|_2^2=o_p\left(m / M_n^1\right), \text{ if } m\log{m /M_n^1} \rightarrow 0.
              \end{equation*}
    \item 
        \begin{equation*}
    \frac{\sqrt{M_n^1}\left(\tilde{\beta}_{j}-\beta^{*}_{j}\right)}{\tilde{\sigma}_{j}} \overset{\mathcal{D}}{\longrightarrow} \mathcal{N}(0,1), \text{ if } m^2\log{m /M_n^1} \rightarrow 0.
              \end{equation*}
\end{enumerate}
\end{thm}

\begin{thm}\label{thm:asy_multi}
    Under an additional partial orthogonality condition that $\{x_j,j \in S\}$ are independent of $\{x_i,i \notin S\}$. Repeat random data splitting for $B$ times and obtain the multi-splitting estimator $\hat{\bbeta}=(\hat{\beta_1}, \dots, \hat{\beta_q})^\top$ defined in~\eqref{smooth}. With known $\boldsymbol{\gamma}$, denote $m=|S|$ and $\hat{\sigma}_j^2=\left(\left\{I^*_{S_{+j}}\right\}^{-1}\right)_{jj}$, for $j \in \{1,\ldots,q\}$. Let $T=\min_i T_i$ and $M_n=\sum_{i=1}^n T_i$. As $n, T ,B \rightarrow \infty$,
\begin{equation*}
    \frac{\sqrt{M_{n}}\left(\hat{\beta}_{j}-\beta_{j}^{*}\right)}{\hat{\sigma}_{j}}\overset{\mathcal{D}}{\longrightarrow}\mathcal{N}(0,1).
    \label{inference}
\end{equation*}
\end{thm}

The theoretical analysis above suggests that the variance of estimating $\hat{\beta}_j$, referred to as $\hat{\sigma}^2_j$, relies on the unknown active set, $S$, while accounting for the data variation across $B$ random splits. Thus, it is impossible to compute $\hat{\sigma}^2_j$ directly in an analytical form. Instead, one can estimate the variance component by employing the infinitesimal jackknife method~\citep{efron2014estimation}, denoted as $\hat{V}_j$. In scenarios where we have obtained a reliable approximation of the estimating variance $\hat{\sigma}^2_j$, for $0<\alpha<1$, the asymptotic $1-\alpha$ confidence interval for $\hat{\beta}_j$ ($j=1, \dots, q$) is given by
\[
\left(\hat{\beta}_j-\Phi^{-1}(1-\alpha / 2) \sqrt{\hat{V}_j}, \hat{\beta}_j+\Phi^{-1}(1-\alpha / 2) \sqrt{\hat{V}_j}\right),
\]
where $\Phi(\cdot)$ is the cumulative distribution function of the standard normal distribution. The $p$-value for testing $\mathrm{H}_0: \beta_j^*=0$ is
\[
2 \times\left\{1-\Phi\left(|\hat{\beta}_j| / \sqrt{\hat{V}_j}\right)\right\}.
\]

The procedure for estimating the variance component of the multi-splitting estimator $\hat{\beta}_j$ is outlined in Algorithm~\ref{alg3}, presented in the Appendix~\ref{appendix:sub:results_variance}. Additionally, a comprehensive derivation procedure for the approximate variance will be provided. It's noteworthy that when $\bgamma$ is estimated, such as the maximum likelihood estimator (MLE) of $\bgamma$, the asymptotic properties of $\hat{\bbeta}_j$ also hold due to the consistency of the profile likelihood estimator.

%%%%%%%%%%%%%%%%%%%%%%%%%%%%%%%%%%%%%%%%%%%%%%%%%%%%%%%%%%

%%%%%%%%%%%%%%%%%%%%%%%%%%%%%%%%%%%%%%%%%%%%%%%%%%%%%%%%%%
% Simulation
%%%%%%%%%%%%%%%%%%%%%%%%%%%%%%%%%%%%%%%%%%%%%%%%%%%%%%%%%%
\section{Simulation}
\label{sec: simulation}
In this section, we evaluate the performance of the proposed approaches for coefficient estimation and inference with high-dimensional covariates and covariance matrix outcomes via simulation studies. 

For $\by_{it}$, data dimension is set to $p=5$ and data are generated from a multivariate normal distribution with mean zero and covariance matrix $\bSigma_{i}$. The covariance matrix has the eigendecomposition of $\bSigma_{i}=\Gamma\Lambda_{i}\Gamma^\top$, where $\Gamma=(\bgamma_{1},\dots,\bgamma_{p})^\top$ is the orthonormal eigenvector matrix assumed to be identical across units and $\Lambda_{i}=\mathrm{diag}\{\lambda_{i1},\dots,\lambda_{ip}\}$ is a diagonal matrix containing $p$ distinct eigenvalues. For the covariates, a dimension of $q=200$ is considered, including the intercept column. $\tilde{\bx}_{i}$'s are generated from a $(q-1)$-dimensional standard normal distribution. Two scenarios are tested in the simulation. (i) The null case, where the eigenvalues, $\lambda_{ij}$, are generated from a log-normal distribution with mean zero and variance of $0.5^2$. (ii) The alternative case, where the second (denoted as C2) and third (denoted as C3) eigenvalues are assumed to satisfy the log-linear model~\eqref{model}. 
That is for these two components, $\lambda_{ij}=\exp \left(\mathbf{x}_i^{\top} \boldsymbol{\beta}\right)$, where $\mathbf{x}_i=(1, \tilde{\bx}_{i}^\top)^{\top}$.
Under the alternative, $\bbeta$ is defined as follows. The intercept term is a random number between $-10$ and $10$. The indices of the active set of C2 is $\{10,20,30\}$ with $\beta$ values $(2,2,-2)$, C3 is $\{15,25, 35\}$ with $\beta$ values $(1,-1,1)$ and those of other components equal to zero except intercept term. Under the above model settings, Assumptions (A1)--(A3) are satisfied. When implementing the proposed approach,  the number of components is determined using the average DfD~\cite[degrees of freedom difference][]{zhao2019covariate} with a threshold of two. The proposed methods employ the asymptotic variance in Theorem \ref{inference}. The multi-splitting procedure introduced in Algorithm~\ref{alg2} is performed with $B=200$ to construct the confidence intervals. All the simulations are repeated 200 times.

We first examine the validity of the proposed method across varying covariate dimensions. We consider two cases: $q=200$ and $q=500$, with a fixed sample size $n=100$ and $T_i=T=100$ for all $i$. Table~\ref{est} displays the performance in estimating and conducting inference on $\bbeta$. In the instance of high-dimensional covariates, the estimated $\hat{\bbeta}$ from the proposed methods closely aligns with the true values in both components. The coverage probability, derived from the asymptotic variance, successfully achieves the designated level of $\alpha = 0.05$ for both components, while keeping the mean squared error (MSE) at a minimal level. It is noteworthy that for particularly large covariate dimensions (e.g., $q = 500$), unsatisfactory results are obtained for C3, while still achieving positive results for C2. To assess the accuracy of estimating $\bgamma$, we use $|\langle\hat{\bgamma},\bgamma\rangle|\rightarrow 1$ as a metric for the estimation error, where $|\langle\cdot,\cdot\rangle|$ represents the inner product of two vectors. Table~\ref{gamma} displays the performance in estimating $\bgamma$. The proposed method yields accurate estimates in both components with a high correlation to the truth in the case where $q=200$. These results collectively indicate that the proposed methods effectively capture the underlying relationships and offer reliable inference in this simulation scenario.

\begin{table}[H]
    \centering
    \caption{\label{est}Estimate and inference on $\bbeta$ in the simulation study. The results are averages over $200$ simulations. Component: corresponding to different eigenvectors; True: real coeffecients; Est.: estimate; SE: standard error; CP: coverage probability; MSE: mean squared error.}
    \label{tab:my-table}
    \begin{tabular}{cccccclcccc}
    \hline
 & \multicolumn{1}{l}{\multirow{2}{*}{Component}} & \multicolumn{4}{c}{$\beta_{10}$} & \multicolumn{1}{l}{}& \multicolumn{4}{c}{$\beta_{25}$} \\ \cline{3-6}  \cline{8-11} 
                       & \multicolumn{1}{l}{} & Truth & Est. (SE)   & CP   & MSE  & & Truth & Est. (SE)    & CP   & MSE  \\ \hline
\multirow{2}{*}{q=200} & C2                   & 2    & 2(0.03)    & 0.97 & 0.01 & & 0    & -0.01(0.03) & 0.98 & 0.02 \\ %\cline{2-10} 
                       & C3                   & 0    & 0.01(0.01) & 0.93 & 0.01 & & -1   & -0.99(0.01) & 0.94 & 0.01 \\ \cline{2-11} 
\multirow{2}{*}{q=500} & C2                   & 2    & 2(0.02)    & 0.98 & 0.06 & & 0    & 0.02(0.02)  & 0.97 & 0.12 \\ %\cline{2-10} 
                       & C3                   & 0    & 1.24(0.01) & 0.11 & 1.79 & & -1   & -0.04(0.01) & 0.05 & 0.97 \\ \cline{1-11} 
\end{tabular}
\end{table}

\begin{table}[H]
 \caption{\label{gamma}Estimate of $\bgamma$ in the simulation study. All values are averaged over 200 simulations. Est.: estimate, SE: standard error}
 \begin{center}
\begin{tabular}{clrrrrr r}
\hline
\multicolumn{1}{l}{} &
  \multicolumn{1}{l}{} &
  \multicolumn{1}{c}{$\gamma_1$} &
  \multicolumn{1}{c}{$\gamma_2$} &
  \multicolumn{1}{c}{$\gamma_3$} &
  \multicolumn{1}{c}{$\gamma_4$} &
  \multicolumn{1}{c}{$\gamma_5$} &
  \multicolumn{1}{c}{$|\langle \hat{\bgamma},\bgamma\rangle|$ (SE)} \\ 
  \hline
\multirow{2}{*}{C2}  & Truth & 0.447 & -0.862 & 0.138  & 0.138 & 0.138 & \multirow{2}{*}{0.816 (0.142)}\\
                                                                             & Est.  & 0.460 & -0.858 & 0.135  & 0.140 & 0.123  \\ \hline
\multirow{2}{*}{C3}  & Truth & 0.447 & 0.138  & -0.862 & 0.138 & 0.138 & \multirow{2}{*}{0.829 (0.126)} \\
                                                                             & Est.  & 0.425 & 0.138  & -0.868 & 0.174 & 0.132  \\ \hline
\end{tabular}
\end{center}
\end{table}

\begin{figure}
  \centering
  \includegraphics[width=1\textwidth]{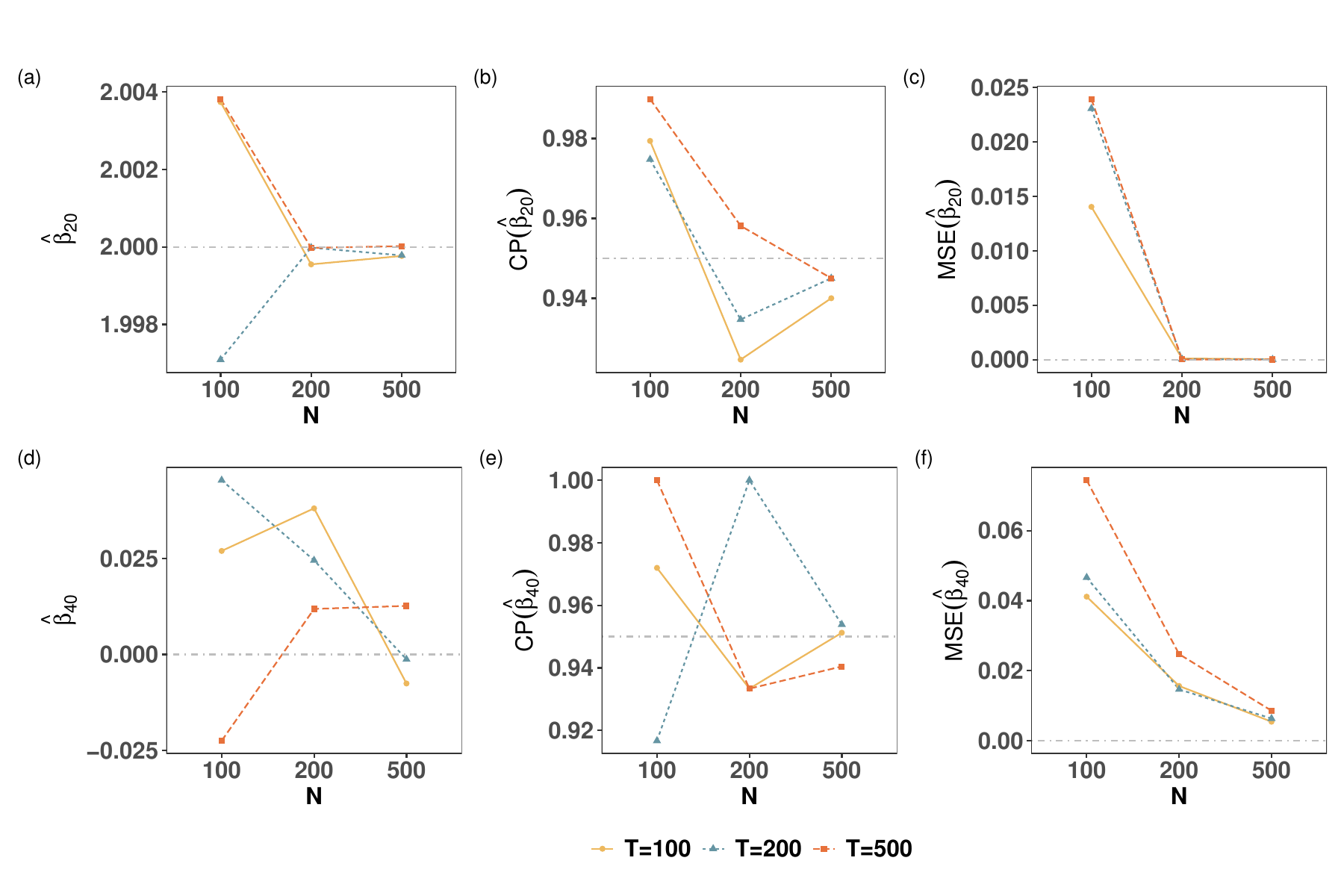}
  \caption{\label{fig:q200}Performance of proposed method. Coverage probability (CP) is derived from the asymptotic variance in the proposed method. The gray dot-dash lines are the true parameters in (a) and (d), the designated level 0.95 in (b) and (e), and zero in (c) and (f). (a) Estimate of $\beta_{20}$ in C2, (b) CP of $\hat{\beta}_{20}$ in C2, (c) MSE of $\hat{\beta}_{20}$ in C2, (d) Estimate of $\beta_{40}$ in C3, (e) CP of $\hat{\beta}_{40}$ in C3, (f) MSE of $\hat{\beta}_{40}$ in C3.}
\end{figure}

To further assess the finite sample performance of the proposed methods, we conduct experiments, varying the number of units ($n = 100, 200, 500$), the number of observations ($T_i = T = 100, 200, 500$), and fixing the dimension of the covariates ($q = 200$). The emphasis is on evaluating the ability to identify nonzero coefficients in each component. The results, displayed in Figure~\ref{fig:q200}, include the estimates, coverage probability derived from the asymptotic variance, and the mean squared error (MSE) of the target model coefficients for both components. The figure illustrates that as both $n$ and $T_i$ increase, the estimates of $\beta_{20}$ in C2 and $\beta_{40}$ in C3 converge to their true values, 2 and 0, respectively. The coverage probability of $\beta_{20}$ in C2 converges to the designated level of 0.95 as both $n$ and $T_i$ increase, while the coverage probability of $\beta_{40}$ in C3 fluctuates around 0.95. The MSEs of estimating $\beta$ in both components converge to zero as $n$ and $T_i$ increase. These findings suggest that the proposed methods demonstrate promising performance in terms of accuracy and reliability as both the sample size and the number of observations increase.

Additionally, we conduct an investigation to assess the impact of changes in the dimensionality of the response variable $\by$ on the performance and present the results in Table~\ref{p_range}. The results illustrate that the proposed method remains remarkably robust to increases in the dimension of $\by$, as long as $p<\min_{i}T_{i}$. This observation emphasizes the reliability and effectiveness of our method across various scenarios.

\begin{table}[H]
    \centering
    \caption{\label{p_range}Estimate and inference on $\bbeta$ as $p$ varies from 5 to 50, while maintaining a fixed sample size of $n=100$ and $T_i=T=100$ for $i=1,\ldots,n$. The results are averages over $200$ simulations. Component: corresponding to different eigenvectors; True: real coeffecients; Est.: estimate; SE: standard error; CP: coverage probability; MSE: mean squared error.}
    \begin{tabular}{cccccclcccc}
    \hline
 & \multicolumn{1}{l}{\multirow{2}{*}{Component}} & \multicolumn{4}{c}{$\beta_{10}$} & \multicolumn{1}{l}{}& \multicolumn{4}{c}{$\beta_{25}$} \\ \cline{3-6}  \cline{8-11} 
                       & \multicolumn{1}{l}{} & Truth & Est. (SE)   & CP   & MSE  & & Truth & Est. (SE)    & CP   & MSE  \\ \hline
\multirow{2}{*}{p=5} & C2                   & 2    & 2(0.03)    & 0.97 & 0.01 & & 0    & -0.01(0.03) & 0.98 & 0.02 \\ %\cline{2-10} 
                       & C3                   & 0    & 0.01(0.01) & 0.93 & 0.01 & & -1   & -0.99(0.01) & 0.94 & 0.01 \\ \cline{2-11} 
\multirow{2}{*}{p=20} & C2                   & 2    & 2(0.03)    & 0.98 & 0.01 & & 0    & 0.02(0.03)  & 0.96 & 0.01 \\ %\cline{2-10} 
                       & C3                   & 0    & 0.02(0.02) & 0.99 & 0.01 & & -1   & -0.95(0.05) & 0.99 & 0.01 \\ \cline{2-11} 
\multirow{2}{*}{p=50} & C2                   & 2    & 2.01(0.05)    & 0.89 & 0.02 & & 0    & -0.02(0.04) & 0.98 & 0.01 \\ %\cline{2-10} 
                       & C3                   & 0    & 0.01(0.02) & 0.99 & 0.01 & & -1   & -1(0.01) & 0.99 & 0.01 \\ \cline{1-11} 
\end{tabular}
\end{table}

%%%%%%%%%%%%%%%%%%%%%%%%%%%%%%%%%%%%%%%%%%%%%%%%%%%%%%%%%%

%%%%%%%%%%%%%%%%%%%%%%%%%%%%%%%%%%%%%%%%%%%%%%%%%%%%%%%%%%
% Real data
%%%%%%%%%%%%%%%%%%%%%%%%%%%%%%%%%%%%%%%%%%%%%%%%%%%%%%%%%%
\section{Application}
\label{sec: application}

We apply the proposed approach to the Lifespan Human Connectome Project (HCP) Aging Study. As an extension of the HCP study of healthy young adults, the Lifespan study utilizes the developed technological advances to offer a high-quality dataset across the human lifespan. The Aging study, in particular, aims to explore a typical aging trajectory and how brain connectome varies among mature and older adults. In this study, the association between two imaging modalities, diffusion tensor imaging (DTI) and resting-state functional magnetic resonance imaging (fMRI), is investigated. DTI is an MRI technique that offers an assessment of brain structural connectivity by tracing the diffusion process of water molecules along white matter fiber bundles. Resting-state fMRI, on the other hand, offers an assessment of the so-called brain functional connectivity, where the fMRI technique measures the blood oxygen level-dependent as a surrogate of neural activity. Hebb's law postulates that frequently communicated brain regions are more likely the consequences of direct structural connections~\citep{hebb2005organization}. Existing literature also suggested that brain structural connectivity shapes and regulates the dynamics of cortical circuits and systems captured by functional connectivity~\citep{sporns2007brain}. Thus, in this study, the output from DTI is considered as the predictor ($X$), and the output from resting-state fMRI is considered as the outcome ($Y$). A total of $n=564$ subjects aged between $36$ and $90$ ($307$ Female and $257$ Male) with both the resting-state fMRI and DTI available without quality control concerns are analyzed. The fMRI data were minimally preprocessed~\citep{glasser2013minimal}. Signals ($\by_{it}$) are extracted from $p=75$ brain regions ($60$ cortical and $15$ subcortical regions) using the Harvard-Oxford atlas in FSL~\citep{smith2004advances} and motion-corrected. The DTI data were preprocessed using developed pipelines for noise reduction, motion, and distortion correction~\cite[more details are described in][]{hall2022tau}. The Desikan atlas~\citep{desikan2006automated} is considered for brain parcellation, which divides the brain into $q=84$ regions of interest (ROIs). Two regions are considered structurally connected if there exists at least one fiber streamline with two endpoints located in these two regions. After building such a network, the degree of each region is summarized and treated as the predictor ($\bx_{i}$). Regions in both the Harvard-Oxford and Desikan atlas are grouped into $10$ functional modules. To better interpret fMRI components, $\bgamma$ is sparsified following an ad hoc procedure using a fused lasso regression~\citep{tibshirani2005sparsity} to impose local smoothness and constancy within each functional module~\citep{grosenick2013interpretable}. 

\begin{figure}
  \begin{center}
    \subfloat[\label{fig:hcp_brain_gamma_C1}C1 rs-fMRI map ($\bgamma$)]{\includegraphics[width=0.33\textwidth]{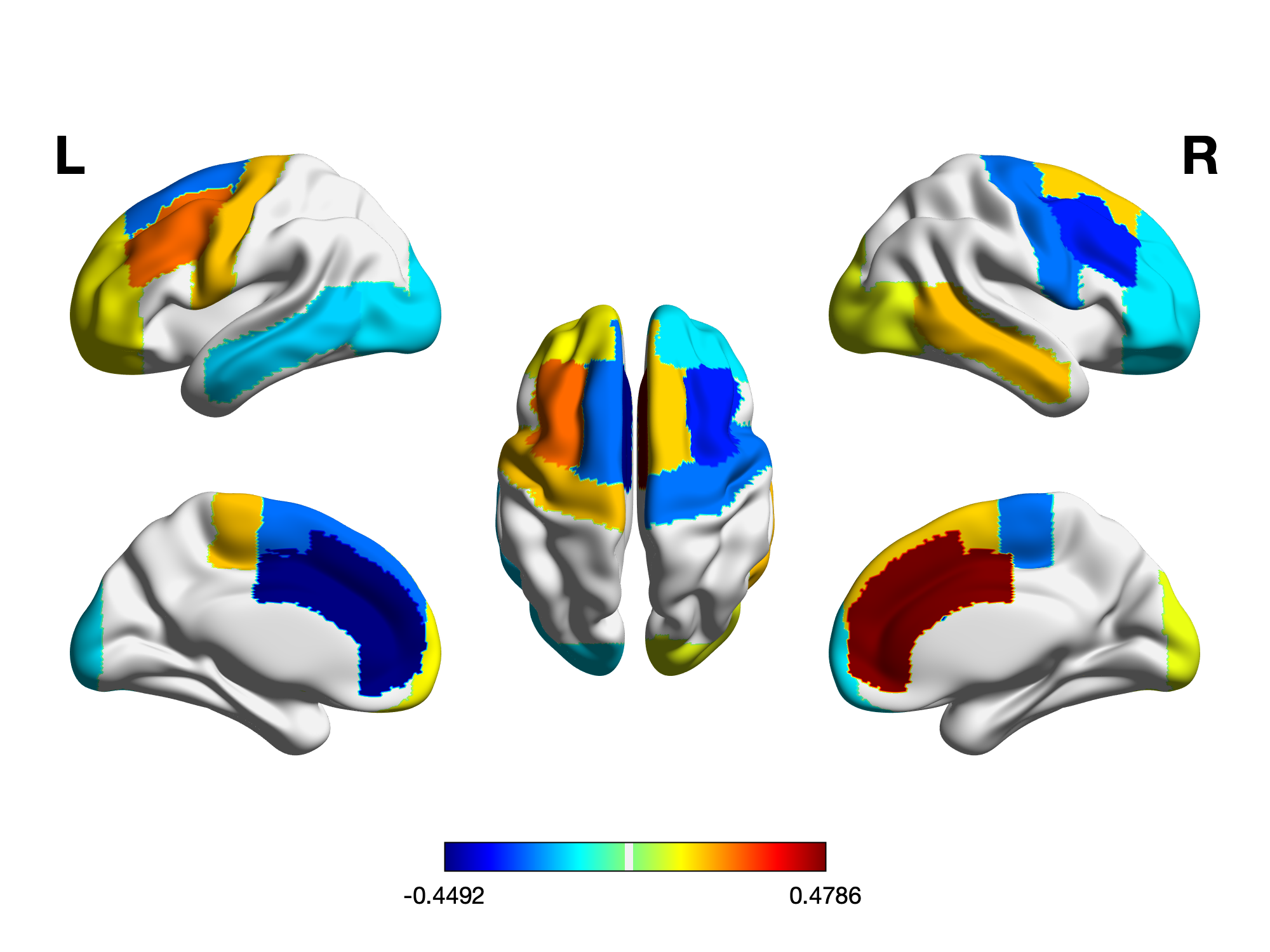}}
    \subfloat[\label{fig:hcp_brain_gamma_C2}C2 rs-fMRI map ($\bgamma$)]{\includegraphics[width=0.33\textwidth]{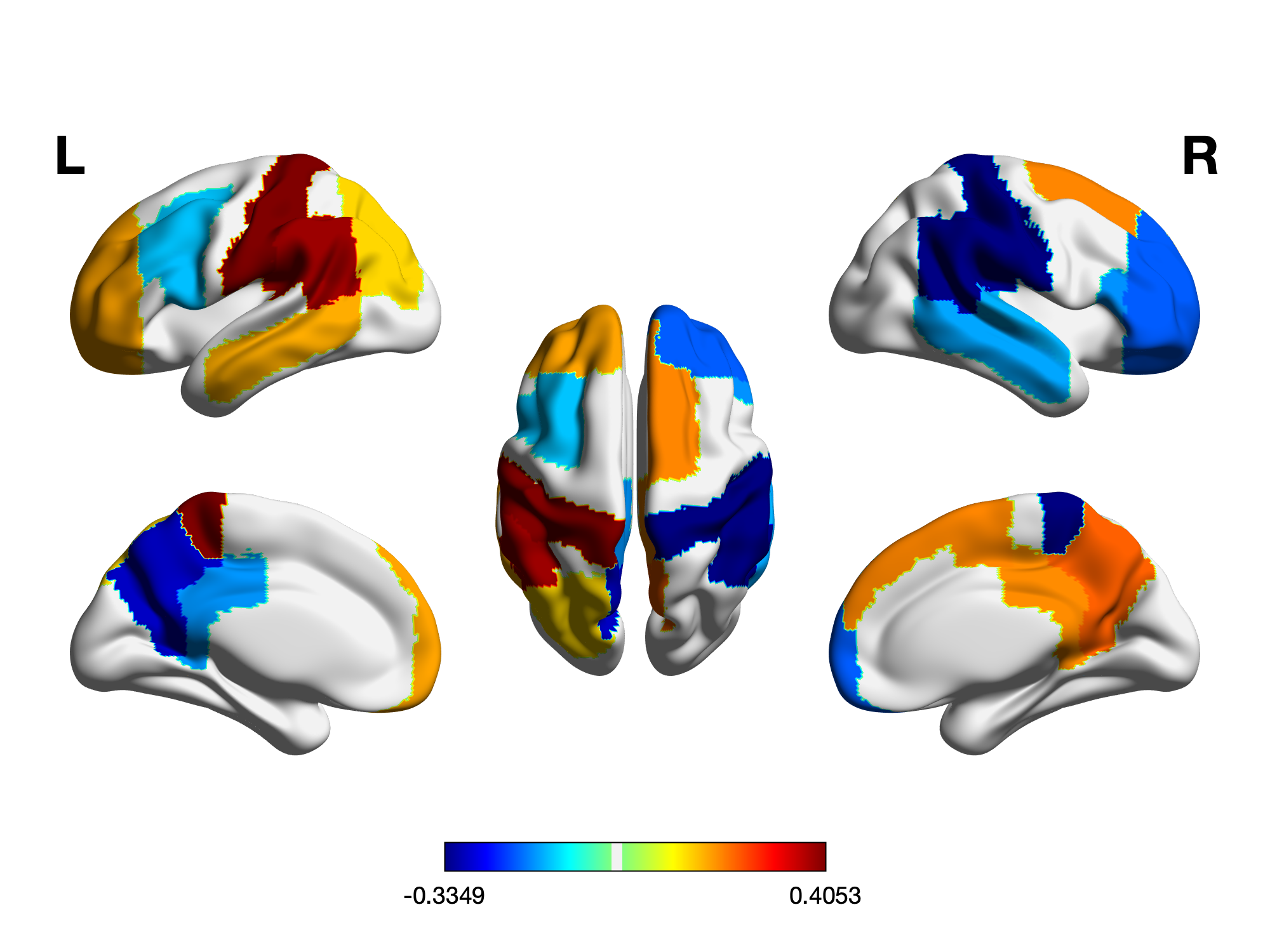}}
    \subfloat[\label{fig:hcp_brain_gamma_C3}C3 rs-fMRI map ($\bgamma$)]{\includegraphics[width=0.33\textwidth]{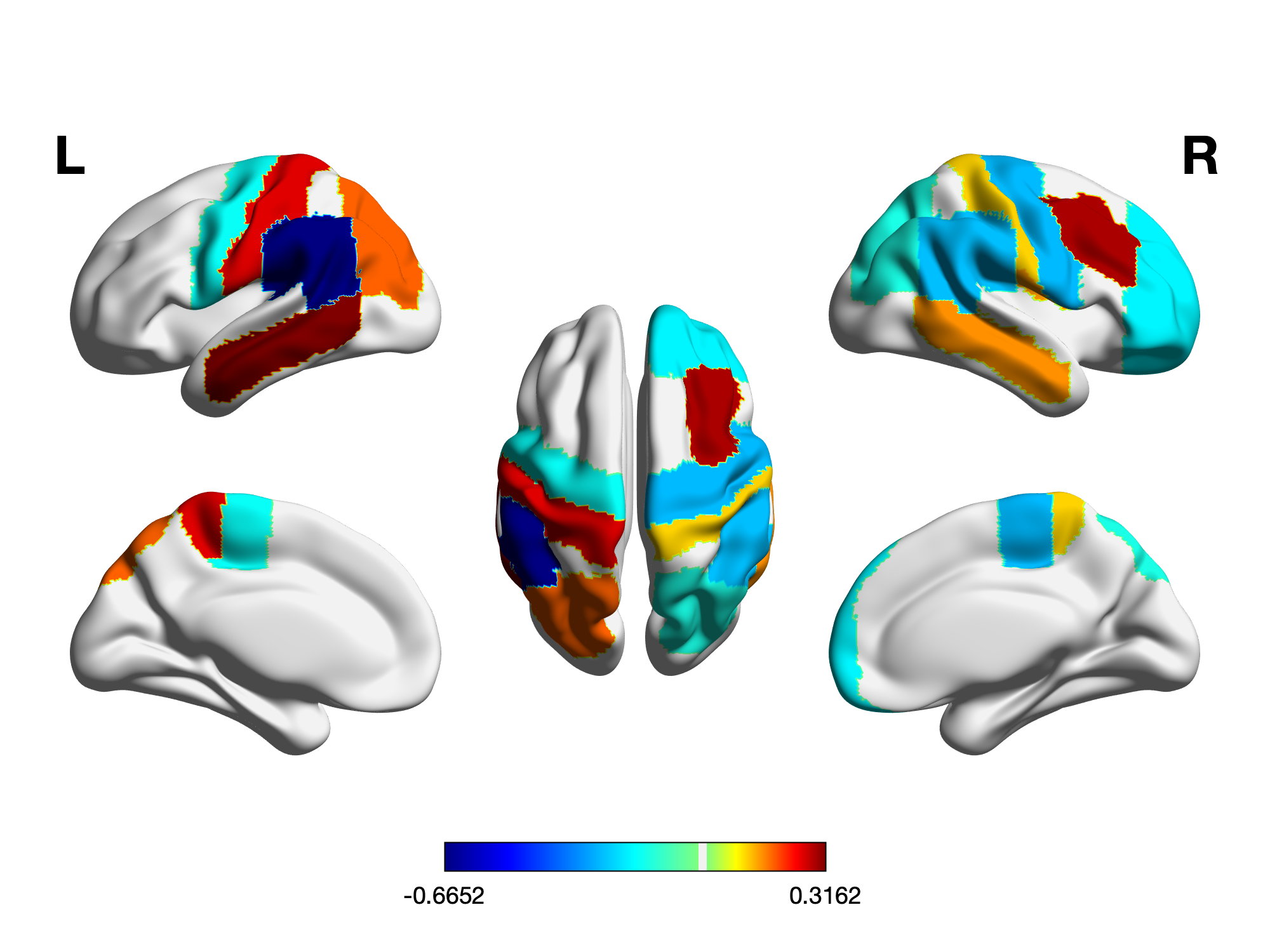}}

    \subfloat[\label{fig:hcp_brain_beta_C1}C1 DTI map ($\beta$)]{\includegraphics[width=0.33\textwidth]{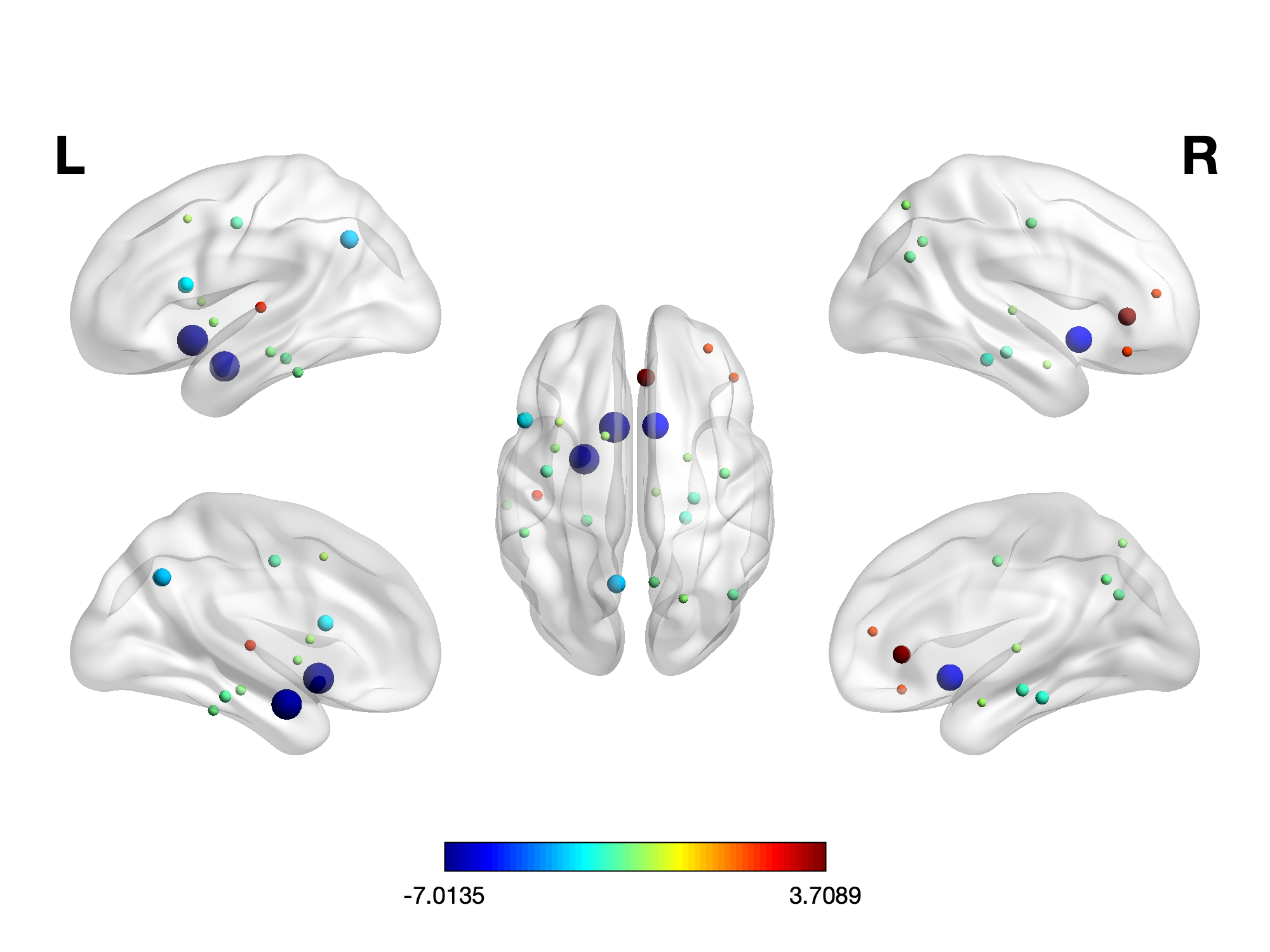}}
    \subfloat[\label{fig:hcp_brain_beta_C2}C2 DTI map ($\beta$)]{\includegraphics[width=0.33\textwidth]{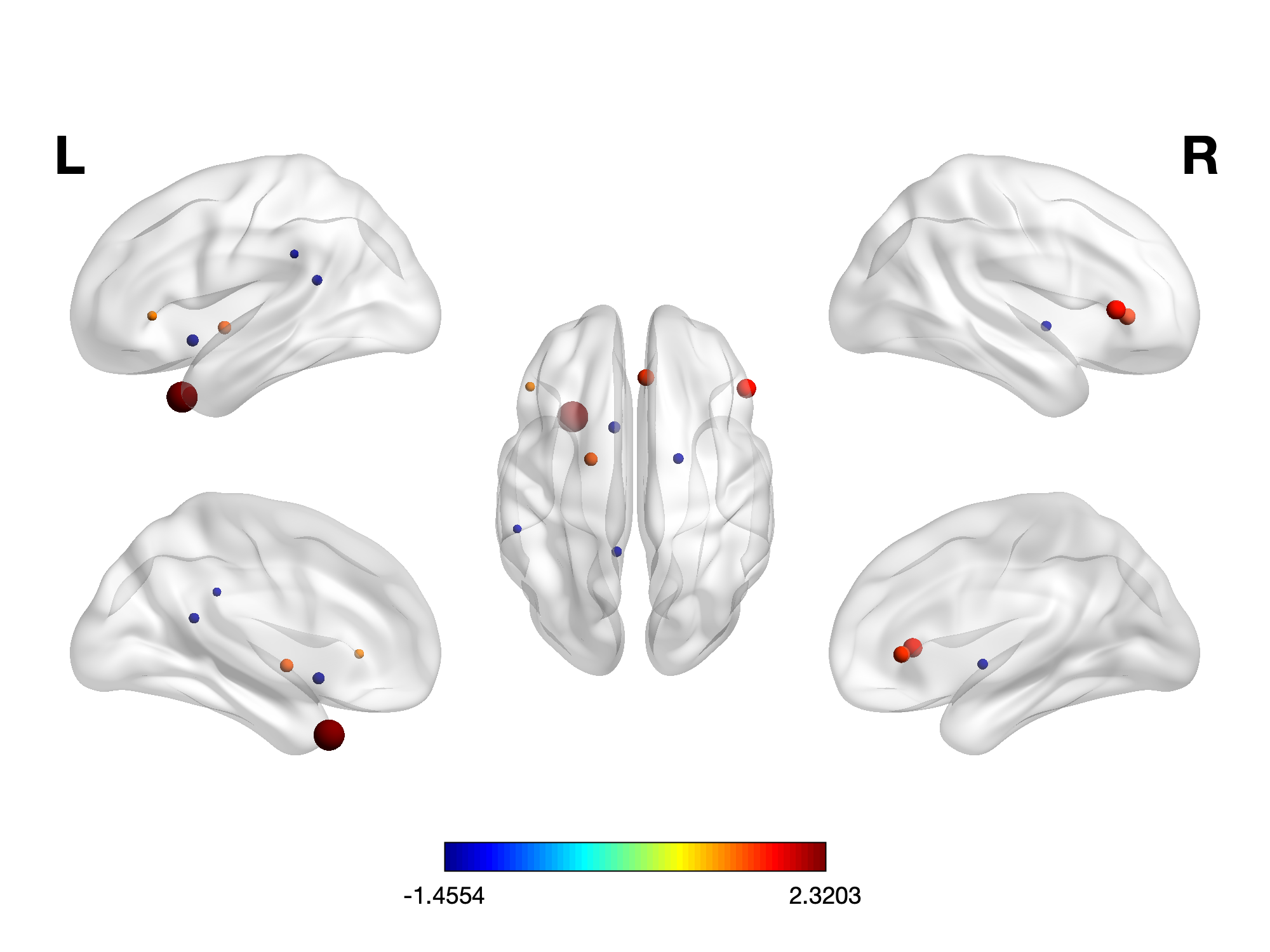}}
    \subfloat[\label{fig:hcp_brain_beta_C3}C3 DTI map ($\beta$)]{\includegraphics[width=0.33\textwidth]{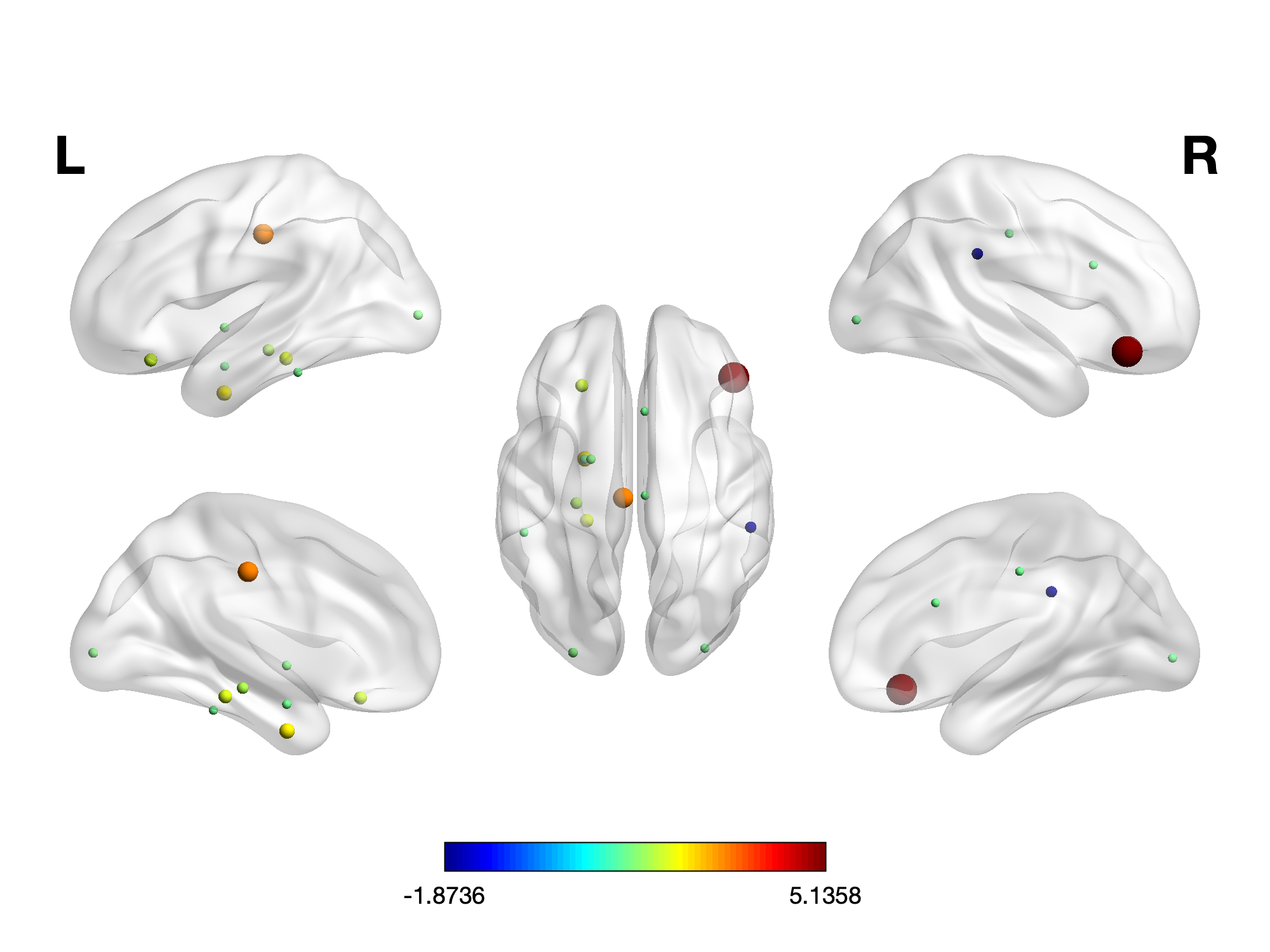}}
  \end{center}
  \caption{\label{fig:hcp_brain}Regions with a nonzero loading in $\bgamma$ from resting-state fMRI and regions with a significant $\beta$ coefficient from DTI in brain maps of the three identified components (C1, C2, and C3) using the proposed approach in the HCP Aging study.}
\end{figure}

Using the $\mathrm{DfD}$ criterion, the proposed approach identifies three orthogonal fMRI components, denoted as C1, C2, and C3. Figure~\ref{fig:hcp_brain} presents the brain maps of regions with a nonzero loading in $\bgamma$ in the resting-state fMRI data, along with the corresponding brain maps of regions with a significant $\beta$ coefficient in the DTI data. The loading profile of all regions is shown in Figure~\ref{fig:hcp_gamma} of the supporting information and the $\beta$ estimate (together with the $95\%$ confidence interval) of all regions is presented in Figure~\ref{fig:hcp_beta}. Using a river plot, configurations by brain functional modules are demonstrated in Figure~\ref{fig:hcp_river}.
A high proportion of overlapping and sign consistency between nonzero $\gamma$'s and significant $\beta$'s is observed in both C1 and C2, consistent with the regulating role of brain white matter integrity on functional connectivity~\citep{jbabdi2011tractography}. In C3, the overlapping region, the supramarginal gyrus, yields the lowest negative value in both $\bgamma$ and $\bbeta$.
A significant portion of the nonzero $\gamma$'s of all three components are regions in the default mode network (DMN), which is more active during the resting state. However, these three components represent three different subnetworks in the DMN. C1 is considered part of the caudal-temporal DMN related to auditory processing and language comprehension~\citep{gollo2018fragility}. C2 is primarily the cingulate-precuneus DMN suggested to play a role in memory and perception~\citep{laird2009investigating}. For C3, the middle temporal gyrus, inferior parietal gyrus, and supramarginal gyrus have been identified as part of a distinct subnetwork within the DMN referred to as the ``temporoparietal junction'' subnetwork playing a role in social cognition, attention, and self-referential processing~\citep{leech2011fractionating}. In summary, the proposed approach identifies brain networks, as well as associated brain structural predictors, that are consistent with existing knowledge of the human brain.

%%%%%%%%%%%%%%%%%%%%%%%%%%%%%%%%%%%%%%%%%%%%%%%%%%%%%%%%%%

%%%%%%%%%%%%%%%%%%%%%%%%%%%%%%%%%%%%%%%%%%%%%%%%%%%%%%%%%%
% Discussion
%%%%%%%%%%%%%%%%%%%%%%%%%%%%%%%%%%%%%%%%%%%%%%%%%%%%%%%%%%
\section{Discussion}
\label{sec: discussion}
In this manuscript, we extend a prior work of covariance regression to the scenario with high-dimensional covariates, where a linear projection on the covariance matrices is assumed and a log-linear model on heteroscedasticity is imposed in the projection space. A regularized estimator of the model coefficient, $\bbeta$, is proposed. Integrated with a likelihood-based optimization criterion, the projection, $\bgamma$, and the model coefficient, $\bbeta$, are identified simultaneously. A novel split-and-smooth procedure is introduced for statistical inference on the estimated model coefficients. Under the assumption of common eigenstructure across covariance matrices, the proposed approach offers asymptotically consistent estimators. Remarkably, our approach stands out for its computational efficiency and estimation reliability requiring fewer assumptions, which is noteworthy as we refrain from making any assumptions regarding the sparsity of the precision matrix of covariates $\bx$.

Several challenges lie ahead for future research. Firstly, theoretical properties when relaxing the common eigenstructure assumption to partial common diagonalization assumption are worth investigating. This is particularly pertinent in real-world scenarios, especially when $p$ is relatively large. Secondly, though asymptotic normality of $\hat{\beta}_j$ is achieved following the proposed inference procedure, the exploration of the covariance structure of the entire $\bbeta$ vector is a promising avenue for future investigation. Thirdly, the present manuscript focuses on the statistical inference of $\bbeta$, future inquiries into the statistical inference of $\bgamma$ will be valuable, particularly in applications, such as fMRI studies, where identifying significant subnetwork structures is crucial. Lastly, considering scenarios where both the response and covariates are high-dimensional, that is both $p$ and $q$ are large, opens up avenues for developing methods and theoretical results tailored to such situations.

%%%%%%%%%%%%%%%%%%%%%%%%%%%%%%%%%%%%%%%%%%%%%%%%%%%%%%%%%%

%%%%%%%%%%%%%%%%%%%%%%%%%%%%%%%%%%%%%%%%%%%%%%%%%%%%%%%%%%
% Acknowledgments
%%%%%%%%%%%%%%%%%%%%%%%%%%%%%%%%%%%%%%%%%%%%%%%%%%%%%%%%%%
% \section*{Acknowledgments}
%%%%%%%%%%%%%%%%%%%%%%%%%%%%%%%%%%%%%%%%%%%%%%%%%%%%%%%%%%

%%%%%%%%%%%%%%%%%%%%%%%%%%%%%%%%%%%%%%%%%%%%%%%%%%%%%%%%%%
% Appendix
%%%%%%%%%%%%%%%%%%%%%%%%%%%%%%%%%%%%%%%%%%%%%%%%%%%%%%%%%%
% \clearpage

\appendix
\counterwithin{figure}{section}
\counterwithin{table}{section}
\counterwithin{equation}{section}
%\counterwithin{lem}{section}
\counterwithin{thm}{section}
\counterwithin{algorithm}{section}

\section*{Appendix}

%========================================================%
This Appendix collects additional theoretical results, the technical proof of the theorems in the main text, and additional data analysis results.
%========================================================%

%========================================================%
% Theory and Proof
%========================================================%
\section{Additional Theoretical Results}
\label{appendix:sec:results}

%========================================================%
\subsection{The variance estimation}
\label{appendix:sub:results_variance}
To define the variance of estimating $\hat{\beta}_j$ in Theorem~\ref{thm:asy_multi}, let $\tilde{\beta}_j^b=\psi(\mathbf{C}_2^b)$ be the estimate given by the $b$th splitting, where $\psi(\cdot)$ is a general function that maps the data to the estimator.
% , thus $\hat{\beta}_j=1/B\sum_{b=1}^B\tilde{\beta}_j^b$. 
Let $N_{ib} \in \{0, 1\}$ indicate whether or not the $i$th unit was sampled in $\mathbf{C}_2^b$. Using the infinitesimal jackknife method, a biased-corrected estimation of the variance of $\hat{\beta}_j$ is introduced as the following. For the $j$th model coefficient,
\begin{equation}
    \hat{V}_j=\frac{n-1}{n}\left(\frac{n}{n-n_1}\right)^2\sum_{i=1}^n\widehat{\operatorname{Cov}}^2_{ij}-\frac{n}{B^2}\frac{n_1}{n-n_1}\sum_{b=1}^B(\tilde{\beta}^b_j-\hat{\beta}_j)^2,
\end{equation}
where
\begin{equation}
    \widehat{\operatorname{Cov}}^2_{ij}=\frac{1}{B}\sum_{b=1}^B\left(N_{ib}-\frac{1}{B}\sum_{b=1}^B N_{ib}\right)\left(\tilde{\beta}_j^b-\hat{\beta}_j\right)
\end{equation}
is the covariance between the estimates, $\tilde{\beta}_j^b$, and the sampling indicators, $N_{ib}$, with respect to the $B$ splits. The initial term of $\hat{V}_j$ is consistent with $Var(\hat{\beta}_j)$ as demonstrated in Theorem 1 of \citet{wager2018estimation}, while the second term of $\hat{V}_j$ is designed to correct bias tailored to the sub-sampling scheme. The process for estimating the variance component of the multi-splitting estimator $\hat{\beta}_j$ is summarized in Algorithm~\ref{alg3}.

\begin{algorithm}
	%\textsl{}\setstretch{1.8}
	\caption{Estimate the variance of the multi-splitting estimator of $\bbeta$.}
	\label{alg3}
	\begin{algorithmic}[1]
		\INPUT 
                   $n_1, n_2$: sample sizes for splitting;
                   
                   $B$: the number of random splits;
                   
                   $\tilde{\boldsymbol{\beta}}^b=\left(\tilde{\beta}_1^b, \dots, \tilde{\beta}_q^b\right)^\top, b=1,2,\ldots,B$;
                   
                   $\hat{\boldsymbol{\beta}}=\left(\hat{\beta}_1, \dots, \hat{\beta}_q\right)^\top$; 

            \For{$j=1, \ldots, q$}
                \State Compute 
                \[
                    \widehat{\operatorname{Cov}}^2_{ij}=\frac{1}{B}\sum_{b=1}^B\left(N_{ib}-N_{i \cdot} \right)\left(\tilde{\beta}_j^b-\hat{\beta}_j\right),
                \]
                 where  $N_{ib} \in \{0, 1\}$ indicate whether or not the $i$th subject was sampled in $\mathbf{C}_2^b$, which is defined above, and $N_{i\cdot} = \frac{1}{B}\sum_{b=1}^B N_{ib}$.
                \State Compute 
                \[
                    \hat{V}_j=\frac{n-1}{n}\left(\frac{n}{n-n_1}\right)^2\sum_{i=1}^n\widehat{\operatorname{Cov}}^2_{ij}-\frac{n}{B^2}\frac{n_1}{n-n_1}\sum_{b=1}^B(\tilde{\beta}^b_j-\hat{\beta}_j)^2.
                \]
            \EndFor
            \State Let $\hat{V}=\left(\hat{V}_1, \dots, \hat{V}_q\right)^\top$.
            \OUTPUT $\hat{V}$.
	\end{algorithmic}  
\end{algorithm}

%========================================================%

%========================================================%
\subsection{Proof of Proposition~\ref{prop:beta_RSC}}
\label{appendix:sub:proof_beta_RSC}
\begin{proof}
The log-likelihood function is:
\begin{equation*}
\begin{aligned}
    \mathcal{L}_n(\beta)&=\frac{1}{2M_n} \sum_{i\in {n}}\left(\mathbf{x}_i^{\top} \boldsymbol{\beta}\right) \cdot T_i+\frac{1}{2M_n} \sum_{i\in {n}} \boldsymbol{\gamma}^{\top} \mathbf{S}_i \boldsymbol{\gamma} \cdot \exp \left(-\mathbf{x}_i^{\top} \boldsymbol{\beta}\right)\\
    &=\frac{1}{M_n}\sum_{i\in {n}}\sum_{t=1}^{T_i} \ell_{it}.\\
    \ell_{i t}^{\prime}(\boldsymbol{\beta})&=-\frac{1}{2} \mathbf{x}_i+\frac{1}{2} \exp \left(-\mathbf{x}_i^{\top} \boldsymbol{\beta}\right)\left(\boldsymbol{\gamma}^{\top} \mathbf{y}_{i t} \mathbf{y}_{i t}^{\top} \boldsymbol{\gamma}\right) \mathbf{x}_i\\
    \ell_{i t}^{\prime \prime}(\boldsymbol{\beta})&=-\frac{1}{2} \exp \left(-\mathbf{x}_i^{\top} \boldsymbol{\beta}\right)\left(\boldsymbol{\gamma}^{\top} \mathbf{y}_{i t} \mathbf{y}_{i t}^{\top} \boldsymbol{\gamma}\right) \mathbf{x}_i \mathbf{x}_i^{\top}
\end{aligned}
\end{equation*}
% $$
% \left(\nabla \mathcal{L}_n(\beta)-\nabla \mathcal{L}_n\left(\beta^*\right)\right)^{\top}\left(\beta-\beta^*\right) \geq \alpha\left\|\beta-\beta^*\right\|_2^2-\tau \frac{\log p}{n}\left\|\beta-\beta^*\right\|_1^2, \quad \forall \beta \in \mathbb{B}_2\left(1 ; \beta^*\right),
% $$
Thus,
$$
\begin{aligned}
\left\langle\nabla \mathcal{L}_n(\beta)-\nabla \mathcal{L}_n\left(\beta^*\right), \beta-\beta^*\right\rangle & =\frac{1}{M_n}\sum_{i\in {n}}\sum_{t=1}^{T_i} \left(\mu\left(\mathbf{x}_i^{\top} \beta\right)-\mu\left(\mathbf{x}_i^{\top} \beta^*\right)\right) \mathbf{x}_i^{\top}\left(\beta-\beta^*\right) \\
& =\frac{1}{M_n}\sum_{i\in {n}}\sum_{t=1}^{T_i} \mu^{\prime}\left(\mathbf{x}_i^{\top} \beta^*+v \mathbf{x}_i^{\top}\left(\beta-\beta^*\right)\right)\left(\mathbf{x}_i^{\top}\left(\beta-\beta^*\right)\right)^2
\end{aligned}
$$
where $\mu\left(\mathbf{x}_i^{\top} \beta\right)=\frac{1}{2} \exp \left(-\mathbf{x}_i^{\top} \boldsymbol{\beta}\right)\left(\boldsymbol{\gamma}^{\top} \mathbf{y}_{i t} \mathbf{y}_{i t}^{\top} \boldsymbol{\gamma}\right)$.

Then from the proof in \cite{negahban2012unified}, there exist positive constants $\kappa_1$ and $\kappa_2$ such that
$$
\left\langle\nabla \mathcal{L}_n(\beta)-\nabla \mathcal{L}_n\left(\beta^*\right), \beta-\beta^*\right\rangle \geq \kappa_1\|\Delta\|_2\left(\|\Delta\|_2-\kappa_2 \sqrt{\frac{\log p}{M_n}}\|\Delta\|_1\right), \quad \forall \beta \in \mathbb{B}_2\left(1 ; \beta^*\right)
$$
with probability at least $1-c_1 \exp \left(-c_2 n\right)$, for some $c_1, c_2>0$. The result follows from the basic arithmetic inequality $2 a b \leq(a+b)^2$.
\end{proof}
%========================================================%

%========================================================%
\subsection{Proof of Theorem~\ref{thm:consistent}}
\label{appendix:sub:proof_consistent}
\begin{proof}
    We assume that the sample fisher matrix $\hat{\Sigma}$ satisfies the RSC condition, and it is not difficult to find that
    \begin{equation*}
        \begin{aligned}
            \hat{\Sigma}&=\frac{1}{2M_n} \sum_{i=1}^n T_i \exp \left(-\mathbf{x}_i^{\top} \boldsymbol{\beta}\right)\left(\boldsymbol{\gamma}^{\top} \mathbf{y}_{i t} \mathbf{y}_{i t}^{\top} \boldsymbol{\gamma}\right) \mathbf{x}_i \mathbf{x}_i^{\top},\\
            \nabla \mathcal{L}_n(\beta)&= -\frac{1}{2M_n} \sum_{i=1}^n T_i \mathbf{x}_i +\frac{1}{2M_n} \sum_{i=1}^n T_i \exp \left(-\mathbf{x}_i^{\top} \boldsymbol{\beta}\right)\left(\boldsymbol{\gamma}^{\top} \mathbf{y}_{i t} \mathbf{y}_{i t}^{\top} \boldsymbol{\gamma}\right) \mathbf{x}_i
        \end{aligned}
    \end{equation*}
    Since data $\{x_i, y_{it}\}$ satisfies the following regression model with a logarithmic link function:
    \begin{equation*}
        \log (\gamma^{\top}\Sigma_i\gamma)=x_i^{\top}\beta
    \end{equation*}
    we can reasonably assume there exist a finite bound that $|\exp \left(-\mathbf{x}_i^{\top} \boldsymbol{\beta}\right)\left(\boldsymbol{\gamma}^{\top} \mathbf{y}_{i t} \mathbf{y}_{i t}^{\top} \boldsymbol{\gamma}\right)| < \xi$.
    
    We can adaptly apply Theorem 3.4 in \cite{lee2015model}, before that, we compute the constants $\kappa_\rho, \kappa_{\varrho}$ and $\kappa_{\mathrm{IC}}$. Since the regularizer is finite (it's a norm), its dual semi-norm is finite. To keep things simple, we let $\varrho=\|\cdot\|_1$. Denote $\mathcal{M}= \operatorname{span}\left(B_{\infty, \mathcal{S}}\right)$, the constant $\kappa_\rho=\kappa_{\varrho}$ is
$$
\kappa_\rho=\sup _\beta\left\{\|\beta\|_1 \mid \beta \in \mathcal{M} \right\}=\sqrt{|\mathcal{S}|} .
$$
Meanwhile, denote $X^T= (\sqrt{T_1}\mathbf{x}_1,\ldots,\sqrt{T_n}\mathbf{x}_n)$, the constant $\kappa_{IC}$ is
\begin{equation*}
    \begin{aligned}
        \kappa_{IC}&= \left\|P_{B_{\infty, \mathcal{S}}}\left(\hat{\Sigma} P_{B_{\infty, \mathcal{S}}}\left(P_{B_{\infty}, \mathcal{S}} \hat{\Sigma} P_{B_{\infty, \mathcal{S}}}\right)^{\dagger} P_{B_{\infty, \mathcal{S}}} z-z\right)\right\|_{\infty} \\
& \quad \leq\left\|X_{\mathcal{S}^c}^T\left(X_{\mathcal{S}}^T\right)^{\dagger} z_{\mathcal{S}}\right\|_{\infty}+\left\|z_{\mathcal{S}^c}\right\|_{\infty} \leq(2-\tau)\|z\|_{\infty}
    \end{aligned}
\end{equation*}

Since the loss function is continuous differentiable, thus any $\lambda<\infty$ satisfies the upper bound in Theorem 3.4 of \cite{lee2015model}. We check our choice also satisfies the lower bound in Theorem 3.4. By \cite{vershynin2010introduction}, Proposition 5.10 and a union bound,
$$
\operatorname{Pr}\left(\left\|\nabla \ell\left(\theta^{\star}\right)\right\|_{\infty}>t\right) \leq \operatorname{Pr}\left(\left\|\sum_{i=1}^n T_i (\frac{1}{2}\xi-\frac{1}{2})\mathbf{x}_i\right\|_{\infty}>M_n t\right) \leq 2 \exp \left(-\frac{M_n t^2}{2 \sigma^2}+\log p\right) .
$$

When $\lambda=\frac{8\xi\kappa_{IC}}{\tau} \sigma \sqrt{\frac{\log p}{M_n}}$, 
\begin{equation*}
\begin{aligned}
& \operatorname{Pr}\left(\frac{4\kappa_{IC}}{\tau}\left\|\nabla \ell\left(\theta^{\star}\right)\right\|_{\infty}>\frac{8(2-\alpha)}{\alpha} \sigma \sqrt{\frac{\log p}{M_n}}\right) \\
& \leq 2 \exp (-2 \log p+\log p)=2 p^{-1} .
\end{aligned}
\end{equation*}
Finally, conclusion is easy to deduce.
\end{proof}
%========================================================%

%========================================================%
\subsection{Proof of Theorem~\ref{thm:consistent_general}}
\begin{proof}
    When $\alpha \neq 0, \mathbf{D}$ is invertible, thus $\mathbf{D}$ has a nontrivial null space. Let $\varrho=\|\cdot\|_1$, the compatibility constants are computed as the following:
$$
\begin{gathered}
\kappa_1=\kappa_{\mathrm{IC}}=\left\|\mathbf{D}_{\mathcal{S}^c} \mathbf{X}^{\top}\left(\tilde{\mathbf{D}}_{\mathcal{S}} \mathbf{X}^{\top}\right)^{-} \operatorname{sign}\left(\boldsymbol{\beta}_{\mathcal{S}}^*\right)\right\|_{\infty}, \\
\kappa_2=\kappa_{\mathcal{R}}=\sup _{\boldsymbol{\beta}}\left\{\|\tilde{\mathbf{D}} \boldsymbol{\beta}\|_1: \boldsymbol{\beta} \in \mathcal{B}_2 \cap \operatorname{span}\left(\tilde{\mathbf{D}}^{\top} \mathcal{B}_{\infty, \mathcal{S}^c}\right)^{\perp}\right\}, \\
\kappa_3=\kappa_{\varrho}=\sup _{\boldsymbol{\beta}}\left\{\|\boldsymbol{\beta}\|_1: \boldsymbol{\beta} \in \mathcal{B}_2 \cap \operatorname{span}\left(\tilde{\mathbf{D}}^{\top} \mathcal{B}_{\infty, \mathcal{S}^c}\right)^{\perp}\right\} .
\end{gathered}
$$
$\mathcal{R}$ and $\varrho$ are finite, $\kappa_1, \kappa_2, \kappa_3<\infty$. The rest of the proof can be found in Lee et al. (2015).
\end{proof}
%========================================================%

%========================================================%
\subsection{Proof of Theorem~\ref{thm:asy_single}}
\begin{proof}
    As a result of the data split, $\mathbf{C}_1$ and $\mathbf{C}_2$ are mutually exclusive, leading to the independence of $\hat{S}$, derived from $\mathbf{C}1$, from $\mathbf{C}2=\left(\mathbf{Y}^2, \mathbf{X}^2\right)$. To analyze the asymptotic properties of single-splitting estimator $\widetilde{\boldsymbol{\beta}}_{\hat{S}{+j}}$ when the number of parameters $m$ diverges, we employ the techniques and findings from He and Shao (2001). Without loss of generality and for the sake of notation simplicity, let us consider the case where $j=1 \in \hat{S}$. The argument holds true if $1 \notin \hat{S}$ or for any other $j$. 

To proceed, we first restrict on the event of $\Omega=\left\{\hat{S} \supseteq S\right\}$. With sure screening property deduced by theorem 1, assume constants $0 \leq c \leq 1/2, K >0$, $\mathrm{P}(\Omega) \geq 1-K\left(p \vee n\right)^{-1-c}$. Recall that
$$
\begin{gathered}
\widetilde{\boldsymbol{\beta}}_{\hat{S}_{+j}}=\underset{\boldsymbol{\beta} \in \mathbf{R}^{|\hat{S}|+1}}{\operatorname{argmin}} \ell_{\hat{S}}\left(\boldsymbol{\beta}_{\hat{S}}\right)=\underset{\boldsymbol{\beta} \in \mathbf{R}^{|\hat{S}|+1}}{\operatorname{argmin}} \ell\left(\boldsymbol{\beta}_{\hat{S}} ; \mathbf{Y}^1, \mathbf{X}_{\hat{S}}^1\right) ; \\
\widetilde{\beta}_1=\left(\widetilde{\boldsymbol{\beta}}_{\hat{S}_{+j}}\right)_1 .
\end{gathered}
$$
Given $M_n$ random samples $(x_1, y_{11}), (x_1, y_{12}), \ldots, (x_1, y_{1 {T_1}}), \ldots, (x_n, y_{n {T_n}})$, the likelihood is
\begin{equation*}
    \mathcal{L}_n(\beta)=\sum_{i=1}^ {n}\sum_{t=1}^ {T_i}\left(\frac{1}{2}\mathbf{x}_i^{\top} \boldsymbol{\beta} -\frac{1}{2}\left(\boldsymbol{\gamma}^{\top} y_{it}y_{it}^{\top} \boldsymbol{\gamma}\right) \cdot \exp \left(-\mathbf{x}_i^{\top} \boldsymbol{\beta}\right)\right)
\end{equation*}
To apply Theorems 2.1 and 2.2 of He and Shao (2000) in our case, we can verify that our Assumptions (A1) and (A2) will lead to their conditions (C1), (C2), (C4) and (C5) with $C=1, r=2$ and $A\left(M_n, m\right)=m$. To verify their (C3), we first note that their $D_n$ is our $I^*_{\hat{S}}$. Then for any $\beta_{\hat{S}}, \alpha \in \mathbf{R}^{m}$ such that $\|\alpha\|_2=1$, a second order Taylor expansion of score function $U_{\hat{S}}\left(\boldsymbol{\beta}_{\hat{S}}\right)$ around $\boldsymbol{\beta}_{\hat{S}}^*$ leads to
$$
\left|\alpha^{\mathrm{T}} \mathbf{E}_{\boldsymbol{\beta}^*}\left(U_{\hat{S}}\left(\boldsymbol{\beta}_{\hat{S}}\right)-U_{\hat{S}}\left(\boldsymbol{\beta}_{\hat{S}}^*\right)\right)-\alpha^{\mathrm{T}} I_{\hat{S}}^*\left(\boldsymbol{\beta}_{\hat{S}}-\boldsymbol{\beta}_{\hat{S}}^*\right)\right| \leq O\left(\left\|\boldsymbol{\beta}_{\hat{S}}-\boldsymbol{\beta}_{\hat{S}}^*\right\|_2^2\right)
$$
Hence,
$$
\sup _{\left\|\boldsymbol{\beta}_{\hat{S}}-\boldsymbol{\beta}_{\hat{S}}^*\right\| \leq\left(m / M_n\right)^{1 / 2}}\left|\alpha^{\mathrm{T}} \mathbf{E}_{\boldsymbol{\beta}^*}\left(U_{\hat{S}}\left(\boldsymbol{\beta}_{\hat{S}}\right)-U_{\hat{S}}\left(\boldsymbol{\beta}_{\hat{S}}^*\right)\right)-\alpha^{\mathrm{T}} I_{\hat{S}}^*\left(\boldsymbol{\beta}_{\hat{S}}-\boldsymbol{\beta}_{\hat{S}}^*\right)\right| \leq O\left(m / M_n\right)=o\left({M_n}^{1 / 2}\right),
$$
which means their (C3) follows. Thus, by Theorem 2.1 of He and Shao (2000),
$$
\left\|\widetilde{\boldsymbol{\beta}}_{\hat{S}}-\boldsymbol{\beta}_{\hat{S}}^*\right\|_2^2=o_p\left(m / M_n^1\right)
$$
given $m \log m / M_n^1 \rightarrow 0$. Furthermore, by Theorem 2.2 of He and Shao (2000), if $m^2 \log m / M_n^1 \rightarrow$ 0, then
$$
\left\|{M_n^1}^{1 / 2}\left(\widetilde{\boldsymbol{\beta}}_{\hat{S}}-\boldsymbol{\beta}_{\hat{S}}^*\right)+{M_n^1}^{-1 / 2}\left\{I_{\hat{S}}^*\right\}^{-1} U_{\hat{S}}\left(\boldsymbol{\beta}_{\hat{S}}^*\right)\right\|_2=o_p(1)
$$
Releasing the restriction on $\Omega$ and with $\mathrm{P}\left(\Omega^c\right)=\mathrm{P}\left({\hat{S}} \nsupseteq S\right) \leq K\left(p \vee n\right)^{-1-c}$, we would still have $\left\|\widetilde{\boldsymbol{\beta}}_{\hat{S}}-\boldsymbol{\beta}_{\hat{S}}^*\right\|_2^2=o_p\left(m / M_n^1\right)$, given $M \log M / m_n^1 \rightarrow 0$. 
% To sce this, for any $\epsilon>0$, we can consider
% $$
% \begin{aligned}
% & \mathrm{P}\left(\left\|\left(n_1 / p_s\right)^{1 / 2}\left(\widetilde{\boldsymbol{\beta}}_S-\boldsymbol{\beta}_S^*\right)\right\|_2>\epsilon\right) \\
% < & \mathrm{P}\left(\left\|\left(n_1 / p_s\right)^{1 / 2}\left(\widetilde{\boldsymbol{\beta}}_S-\boldsymbol{\beta}_S^*\right)\right\|_2>\epsilon \mid \Omega\right) \mathrm{P}(\Omega)+\mathrm{P}\left(\Omega^c\right) \\
% < & \mathrm{P}\left(\left\|\left(n_1 / p_s\right)^{1 / 2}\left(\widetilde{\boldsymbol{\beta}}_S-\boldsymbol{\beta}_S^*\right)\right\|_2>\epsilon \mid \Omega\right)+K_2\left(p \vee n_2\right)^{-1-c_2}
% \end{aligned}
% $$
% where both terms in the last inequality converge to 0 as $n_1 \rightarrow \infty$ and $n_2=(1-q) n_1 / q$, with $0<q<1$ a constant. 

Similarly, we can show $\left\|{M_n^1}^{1 / 2}\left(\widetilde{\boldsymbol{\beta}}_{\hat{S}}-\boldsymbol{\beta}_{\hat{S}}^*\right)+{M_n^1}^{-1 / 2}\left\{I_{\hat{S}}^*\right\}^{-1} U_{\hat{S}}\left(\boldsymbol{\beta}_{\hat{S}}^*\right)\right\|_2=$ $o_p(1)$, if $m^2 \log m / M_n^1 \rightarrow 0$, which can also be written as
$$
\widetilde{\boldsymbol{\beta}}_{\hat{S}}-\boldsymbol{\beta}_{\hat{S}}^*=-{M_n^1}^{-1}\left\{I_{\hat{S}}^*\right\}^{-1} U_{\hat{S}}\left(\boldsymbol{\beta}_{\hat{S}}^*\right)+r_{M_n^1}
$$
with $\left\|r_{M_n^1}\right\|_2^2=o_p\left(1 / M_n^1\right)$. Consequently, by taking $\alpha=(0,1,0, \ldots, 0)^{\mathrm{T}}$ and left-multiplying both sides by ${M_n^1}^{1 / 2} \alpha^{\mathrm{T}}$, we have
$$
\sqrt{M_n^1}\left(\widetilde{\beta}_1-\beta_1^*\right) / \widetilde{\sigma}_1 \stackrel{d}{\rightarrow} N(0,1),
$$
where $\widetilde{\sigma}_1^2=\left(\left\{I_{\hat{S}}^*\right\}^{-1}\right)_{11}$.
\end{proof}
%========================================================%

%========================================================%
\subsection{Proof of Theorem~\ref{thm:asy_multi}}
Similar to Lemma 4 in \cite{fei2021estimation}, we apply the lemma here to our case without proof.
\begin{lem}
\label{lemma}
    With model and Assumptions (A1) and (A2), consider the estimator $\widetilde{\boldsymbol{\beta}}_{\hat{S}}$ with respect to subset ${\hat{S}}$ as defined above. Denote by $m=|{\hat{S}}|$. If $m \log m / M_n \rightarrow 0$, then with probability going to $1,\left|\widetilde{\boldsymbol{\beta}}_{\hat{S}}\right|_{\infty} \leq C_\beta$, where $C_\beta>0$ is a constant depending on $c_{\min }, c_{\max }, c_\beta.$
\end{lem}

Now we use the Lemma to prove the result of Theorem~\ref{thm:asy_multi}.
\begin{proof}
We define the oracle estimators of $\beta_j^*$ on the full data $(\mathbf{Y}, \mathbf{X})$ and the $b$-th subsample $\mathbf{C}_2^b$ respectively, where the candidate set is the true set $S$ and $|S|=s_0$:
$$
\begin{aligned}
& \check{\boldsymbol{\beta}}_{S_{+j}}=\underset{\boldsymbol{\beta} \in \mathbf{R}^{s_0+1}}{\operatorname{argmin}} \ell_{S_{+j}}\left(\boldsymbol{\beta}_{S_{+j}}\right)=\underset{\boldsymbol{\beta} \in \mathbf{R}^{s_0+1}}{\operatorname{argmin}} \ell_{S_{+j}}\left(\boldsymbol{\beta}_{S_{+j}} ; \mathbf{Y}, \mathbf{X}_{S_{+j}}\right), \check{\boldsymbol{\beta}}_j=\left(\check{\boldsymbol{\beta}}_{S_{+j}}\right)_j ; \\
& \check{\boldsymbol{\beta}}_{S_{+j}}^b=\underset{\boldsymbol{\beta} \in \mathbf{R}^{s_0+1}}{\operatorname{argmin}} \ell_{S_{+j}}^b\left(\boldsymbol{\beta}_{S_{+j}}\right)=\underset{\boldsymbol{\beta} \in \mathbf{R}^{s_0+1}}{\operatorname{argmin}} \ell_{S_{+j}}\left(\boldsymbol{\beta}_{S_{+j}} ; \mathbf{Y}^{1(b)}, \mathbf{X}_{S_{+j}}^{1(b)}\right), \check{\boldsymbol{\beta}}_j^b=\left(\check{\boldsymbol{\beta}}_{S_{+j}}^b\right)_j . \\
&
\end{aligned}
$$
By Theorem~\ref{thm:asy_single} and given $s_0^2 \log s_0 / n \rightarrow 0$, for each $j \in\{1, \ldots, p\}$,
\begin{equation}
    \sqrt{M_n^1}\left(\check{\beta}_j-\beta_j^*\right) / \check{\sigma}_j \stackrel{d}{\rightarrow} N(0,1) \quad \text { as } M_n^1 \rightarrow \infty,
    \label{asynormal}
\end{equation}
where $\check{\sigma}_j^2=\left(\left\{I_{S_{+j}}^*\right\}^{-1}\right)_{j j}$.
With the oracle estimators $\breve{\beta}_j$ 's and $\breve{\beta}_j^b$,s, we have the following decomposition:
$$
\begin{aligned}
& \sqrt{M_n}\left(\widehat{\beta}_j-\beta_j^*\right) \\
= & \sqrt{M_n}\left(\check{\beta}_j-\beta_j^*\right)+\sqrt{M_n}\left(\widehat{\beta}_j-\check{\beta}_j\right) \\
= & \sqrt{M_n}\left(\check{\beta}_j-\beta_j^*\right)+\sqrt{M_n}\left(\frac{1}{B} \sum_{b=1}^B \widetilde{\beta}_j^b-\check{\beta}_j\right) \\
= & \underbrace{\sqrt{M_n}\left(\check{\beta}_j-\beta_j^*\right)}_{\mathrm{I}}+\underbrace{\sqrt{M_n}\left(\frac{1}{B} \sum_{b=1}^B \check{\beta}_j^b-\check{\beta}_j\right)}_{\mathrm{II}}+\underbrace{\sqrt{M_n}\left(\frac{1}{B} \sum_{b=1}^B\left\{\widetilde{\beta}_j^b-\check{\beta}_j^b\right\}\right)}_{\mathrm{III}} .
\end{aligned}
$$
The initial two terms in the aforementioned decomposition, which do not encompass various selections $S^b$'s, pertaining to the oracle estimators and the true active set $S$. We need to show the following, as it will subsequently yield the outcomes articulated in the theorem through the application of Slutsky's theorem.
\begin{itemize}
    \item[(a)] $\mathrm{I} / \check{\sigma}_j=\sqrt{M_n}\left(\check{\beta}_j-\beta_j^*\right) / \check{\sigma}_j \stackrel{d}{\rightarrow} N(0,1)$; 
    \item[(b)] $\mathrm{II}=\frac{\sqrt{M_n}}{B} \sum_{b=1}^B\left\{\check{\beta}_j^b-\check{\beta}_j\right\}=o_p(1)$;
    \item[(c)] $\mathrm{III}=\frac{\sqrt{M_n}}{B} \sum_{b=1}^B\left\{\widetilde{\beta}_j^b-\check{\beta}_j^b\right\}=o_p(1)$.]
\end{itemize}

First, (a) holds because of \eqref{asynormal}. To show (b), i.e. $\mathrm{II}=o_p(1)$, we first denote $\xi_{b, M_n}=$ $\sqrt{M_n}\left(\check{\beta}_j^b-\check{\beta}_j\right)$, then II $=\left(\sum_{b=1}^B \xi_{b, M_n}\right) / B$. Since the sampling indicator vectors, $N_{ib}$ 's (defined in above Appendix~\ref{appendix:sub:results_variance}) are i.i.d, $\xi_{b, M_n}$ 's are i.i.d conditional on data $\mathbf{C}^{(n)}=(\mathbf{Y}, \mathbf{X})$. The conditional distribution of $\sqrt{M_n}\left(\check{\beta}_j^b-\check{\beta}_j\right)$ given $\mathbf{C}^{(n)}$ is asymptotically the same as the unconditional distribution of $\sqrt{M_n}\left(\check{\beta}_j-\beta_j^*\right)$, which converges to zero Gaussian by \eqref{asynormal}. With the uniform boundedness of $\check{\beta}_j^b$ and $\check{\beta}_j$ as shown in Lemma~\ref{lemma}, we can show that $\mathbf{E}\left(\xi_{b, M_n} \mid \mathbf{C}^{(n)}\right) \rightarrow 0$ and $\operatorname{Var}\left(\xi_{b, M_n} \mid \mathbf{C}^{(n)}\right) \rightarrow \check{\sigma}_j^2$ uniformly over $\mathbf{C}^{(n)}$ as $M_n \rightarrow \infty$. Furthermore, $\mathbf{E}\left(\mathrm{II} \mid \mathbf{C}^{(n)}\right)=$ $\mathbf{E}\left(\xi_{b, M_n} \mid \mathbf{C}^{(n)}\right)$, and $\operatorname{Var}\left(\mathrm{II} \mid \mathbf{C}^{(n)}\right)=\operatorname{Var}\left(\xi_{b, M_n} \mid \mathbf{C}^{(n)}\right) / B$. By Chebyshev Inequality, for any $\delta, \zeta>0$, there exist $N_0, B_0>0$ such that when $n>N_0, B>B_0$, $\mathrm{P}(|\mathrm{II}| \geq \delta) \leq \zeta$. Thus, $\mathrm{II}=o_p(1)$.
% $$
% \begin{aligned}
% & \mathrm{P}(|\mathrm{II}| \geq \delta) \leq \int_{\Omega_n} \mathrm{P}\left(|\mathrm{II}| \geq \delta \mid \mathbf{D}^{(n)}\right) \mathrm{dP}\left(\mathbf{D}^{(n)}\right) \\
% \leq & \int_{\Omega_n} \mathrm{P}\left(\left|\mathrm{II}-\mathbf{E}\left(\mathrm{II} \mid \mathbf{D}^{(n)}\right)\right| \geq \delta / 2 \mid \mathbf{D}^{(n)}\right) \mathrm{dP}\left(\mathbf{D}^{(n)}\right) \\
% \leq & \int_{\Omega_n} \frac{\operatorname{Var}\left(\mathrm{II} \mid \mathbf{D}^{(n)}\right)}{\delta^2 / 4} \mathrm{dP}\left(\mathbf{D}^{(n)}\right) \leq \frac{\check{\sigma}_j^2}{B_0 \delta^2 / 4} \int_{\Omega_n} \mathrm{dP}\left(\mathbf{D}^{(n)}\right) \leq \zeta .
% \end{aligned}
% $$

To prove (c), i.e. III $=o_p(1)$, we first note that each subsample $\mathbf{C}_1^b$ can be regarded as a random sample of $n_1=k n(0<k<1)$ i.i.d. observations from the population distribution for which Theorem~\ref{thm:consistent} holds, that is to say $\left|\hat{S}^b\right| \leq K_1 n^{c_1}$ and $\mathrm{P}\left(S \subseteq \hat{S}^b\right) \geq 1-K_2(q \vee n)^{-1-c_2}$. We show that for any $b$, conditional on $\hat{S}^b \supseteq S, \sqrt{M_n}\left(\widetilde{\beta}_j^b-\check{\beta}_j^b\right)=o_p(1)$.

To see this, we first arrange the order of the components of $\mathbf{x}=\left(x_1, \ldots, x_q\right)$ such that the first $s_0$ components are signal variables. In other words, $S=\left\{1, \ldots, s_0\right\}$. From the proof of Theorem~\ref{thm:asy_single} and omitting superscript $b$, we have that
\begin{equation}\label{eq:residual}
    \begin{aligned}
& \widetilde{\beta}_j-\beta_j^*=-{M_n^1}^{-1} \widetilde{e}_j^{\mathrm{T}}\left\{I_{S_{+j}}^*\right\}^{-1} U_{\hat{S}_{+j}}\left(\boldsymbol{\beta}_{\hat{S}_{+j}}^*\right)+\widetilde{r}_{M_n^1}, \\
& \check{\beta}_j-\beta_j^*=-{M_n^1}^{-1} \check{e}_j^{\mathrm{T}}\left\{I_{S_{+j}^*}^*\right\}^{-1} U_{S_{+j}}\left(\boldsymbol{\beta}_{S_{+j}}^*\right)+\check{r}_{M_n^1},
\end{aligned}
\end{equation}
where $\widetilde{e}_j=(0, \ldots, 0,1,0, \ldots, 0)^{\mathrm{T}}$ is a unit vector of length $\left|\hat{S}_{+j}\right|$ to index the position of variable $j$ in $\hat{S}_{+j}$, $\check{e}_j$ is a unit vector of length $\left|S_{+j}\right|$ to index the position of variable $j$ in $S_{+j}$, and the residuals $\left\|\widetilde{r}_{M_n^1}\right\|_2^2=o_p\left(1 / M_n^1\right),\left\|\check{r}_{M_n^1}\right\|_2^2=o_p\left(1 / n_1\right)$. Here, $I_{\hat{S}_{+j}}^*$ and $I_{S_{+j}}^*$ are two submatrices of the expected information at $\boldsymbol{\beta}^*$, which is derived in the proof of Proposition~\ref{prop:beta_RSC} as $I^* = \mathbb{E}\left\{\frac{1}{2M_n}\sum_{i\in {n}}\sum_{t=1}^{T_i} \exp \left(-\mathbf{x}_i^{\top} \boldsymbol{\beta}\right)\left(\boldsymbol{\gamma}^{\top} \mathbf{y}_{i t} \mathbf{y}_{i t}^{\top} \boldsymbol{\gamma}\right) \mathbf{x}_i \mathbf{x}_i^{\top}\right\} = \mathbb{E}\left\{\frac{1}{n} \mathbf{X}^{\mathrm{T}} \mathbf{V} \mathbf{X}\right\}$, where $\mathbf{V}=\operatorname{diag}\left\{\nu_1, \ldots, \nu_n\right\}$ is an $n \times n$ diagonal matrix with $\nu_i=\frac{1}{2T_i}\sum_{t=1}^{T_i} \exp \left(-\mathbf{x}_i^{\top} \boldsymbol{\beta}\right)\left(\boldsymbol{\gamma}^{\top} \mathbf{y}_{i t}\right.$ $ \left. \mathbf{y}_{i t}^{\top} \boldsymbol{\gamma}\right)$.

For any $j \in S, k \in S^c$, the complement of $S$, the partial orthogonality condition \citep{fan2008sure} that $\left\{x_j, j \in S\right\}$ are independent of $\left\{x_k, k \in S^c\right\}$ implies that $I^*$ is block-diagonal with two blocks indexed by $S$ and $S^c$. That is,
$$
I^*=\left(\begin{array}{cc}
\mathbf{E}\left(\frac{1}{n} \mathbf{X}_{S}^{\mathrm{T}} \mathbf{V} \mathbf{X}_{S}\right) & 0 \\
0 & \mathbf{E}\left(\frac{1}{n} \mathbf{X}_{S^c}^{\mathrm{T}} \mathbf{V} \mathbf{X}_{S^c}\right)
\end{array}\right) .
$$
where the submatrices $\mathbf{X}_{S}$ and $\mathbf{X}_{S^c}$ are submatrices of $\mathbf{X}$. Hence, $I_{\hat{S}_{+j}}^*$ is blockdiagonal with two blocks indexed by $S$ and $\hat{S}_{+j} \backslash S$, and $I_{S_{+j}}^*$ is block-diagonal with two blocks indexed by $S$ and $S_{+j} \backslash S=\emptyset$ if $j \in S$ or $=\{j\}$ otherwise.

Therefore, $\left\{I^*\right\}^{-1},\left\{I_{\hat{S}_{+j}}^*\right\}^{-1}$ and $\left\{I_{S_{+j}}^*\right\}^{-1}$ are all block-diagonal. Write $U\left(\boldsymbol{\beta}^*\right)=\left(u_0, u_1, \ldots, u_p\right)^{\mathrm{T}}$,
$\widetilde{e}_j^{\mathrm{T}}\left\{I_{\hat{S}_{+j}}^*\right\}^{-1}=\left(\widetilde{i}_{j k}\right)_{k \in \hat{S}_{+j}}$ and $\check{e}_j^{\mathrm{T}}\left\{I_{S_{+j}}^*\right\}^{-1}=\left(\check{i}_{j k}\right)_{k \in S_{+j}}$. Then, it follows that $\widetilde{i}_{j k}=\check{i}_{j k}$ for $k \in S$, which leads to
$$
\begin{aligned}
\sqrt{M_n^1}\left(\widetilde{\beta}_j-\check{\beta}_j\right) & =-\frac{1}{\sqrt{M_n^1}} \sum_{k \in \hat{S}_{+j}} \widetilde{i}_{j k} u_k+\frac{1}{\sqrt{M_n^1}} \sum_{k \in S_{+j}} \check{i}_{j k} u_k+r_{n_1}^{\prime} \\
& =-\frac{1}{\sqrt{M_n^1}} \sum_{k \in \hat{S} \backslash S} \widetilde{i}_{j k} u_k+\frac{1\left(j \notin S\right)}{\sqrt{M_n^1}}\left(\check{i}_{j j}-\widetilde{i}_{j j}\right) u_j+r_{n_1}^{\prime}
\end{aligned}
$$
where $r_{n_1}^{\prime}=\sqrt{n_1}\left(\widetilde{r}_{n_1}-\check{r}_{n_1}\right)=o_p(1)$, and $\widetilde{r}_{n_1}$ and $\check{r}_{n_1}$ are as in~\eqref{eq:residual}.
As $\left|\hat{S} \backslash S\right| \leq K_1 n^{c_1}=o\left(\sqrt{n_1}\right)$ with $0 \leq c_1<1 / 2$, $\operatorname{Var}\left(\sum_{k \in \hat{S} \backslash S} \widetilde{i}_{j k} u_k\right)=o\left(n_1\right)$. By the Chebyshev inequality, the first term on the right-hand side converges to 0 in probability. Thus, each of these three terms is $o_p(1)$ and we have $\sqrt{M_n^1}\left(\widetilde{\beta}_j-\check{\beta}_j\right)=o_p(1)$. As $n_1 / n=k$ where $0<k<1$, the original statement holds.

Now define $\eta_b=\mathbbm{1}\left\{S \nsubseteq \hat{S}^b\right\} \sqrt{M_n}\left(\widetilde{\beta}_j^b-\check{\beta}_j^b\right)$, while omitting subscripts $j$ in $\eta$ for simplicity, then III $=\left(\sum_{b=1}^B \eta_b\right) / B$. When $S \nsubseteq \hat{S}^b$, by Lemma~\ref{lemma}, there exists a $C_\beta \geq c_\beta$ such that $\sup _b\left|\widetilde{\beta}_j^b-\check{\beta}_j^b\right| \leq \sup _b\left|\widetilde{\beta}_j^b-\beta_j^*\right|+\sup _b\left|\check{\beta}_j^b-\beta_j^*\right| \leq 2 C_\beta+1$. Therefore,
\begin{equation}\label{eq:var}
    \begin{aligned}
\mathbb{E}\left(\eta_b\right) & \leq \mathrm{P}\left(S \nsubseteq \hat{S}^b\right) \sqrt{M_n} \sup _{1 \leq b \leq B}\left|\widetilde{\beta}_j^b-\check{\beta}_j^b\right| \leq 2 C_\beta \sqrt{M_n} K_2(q \vee n)^{-1-c_2}, \\
\operatorname{Var}\left(\eta_b\right) & \leq \mathrm{P}\left(S \nsubseteq \hat{S}^b\right) M_n \sup _{1 \leq b \leq B}\left(\widetilde{\beta}_j^b-\check{\beta}_j^b\right)^2 \leq 4 C_\beta^2 M_n K_2(q \vee n)^{-1-c_2} .
\end{aligned}
\end{equation}
With dependent $\eta_b$ 's, we further have
$$
\begin{gathered}
\mathbf{E}(\mathrm{III})=\mathbf{E}\left\{\left(\sum_{b=1}^B \eta_b\right) / B\right\} \leq 2 C_\beta \sqrt{M_n} K_2(q \vee n)^{-1-c_2}, \\
\operatorname{Var}(\mathrm{III}) \leq \frac{1}{B^2} \sum_{b=1}^B \sum_{b^{\prime}=1}^B\left|\operatorname{Cov}\left(\eta_b, \eta_{b^{\prime}}\right)\right| \leq 4 C_\beta^2 M_n K_2(q \vee n)^{-1-c_2},
\end{gathered}
$$
where the last inequality holds because of $\left|\operatorname{Cov}\left(\eta_b, \eta_{b^{\prime}}\right)\right| \leq\left\{\operatorname{Var}\left(\eta_b\right) \operatorname{Var}\left(\eta_{b^{\prime}}\right)\right\}^{1 / 2}$ and~\eqref{eq:var}. Then we can show III $=o_p(1)$. 
\end{proof}
%========================================================%

%========================================================%
% Additional Application Results
%========================================================%
\section{Additional Application Results}
\label{appendix:sec:application}

\begin{figure}[h]
  \begin{center}
    \subfloat[\label{fig:hcp_gamma_C1}C1]{\includegraphics[width=0.81\textwidth]{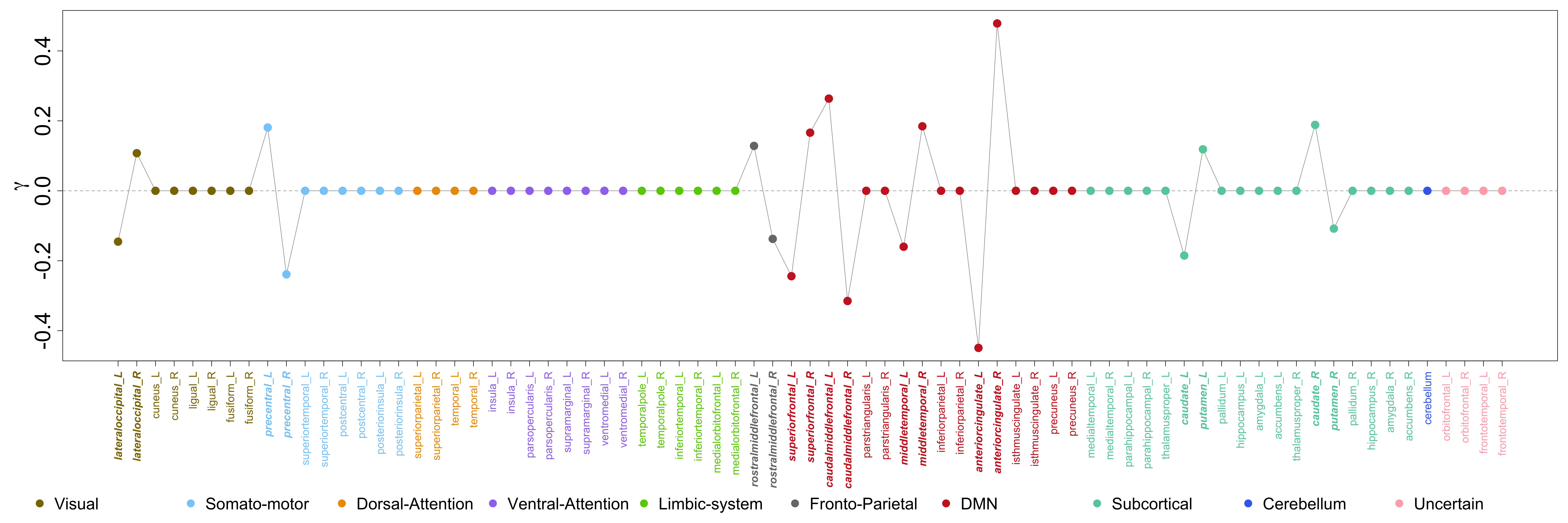}}

    \subfloat[\label{fig:hcp_gamma_C2}C2]{\includegraphics[width=0.81\textwidth]{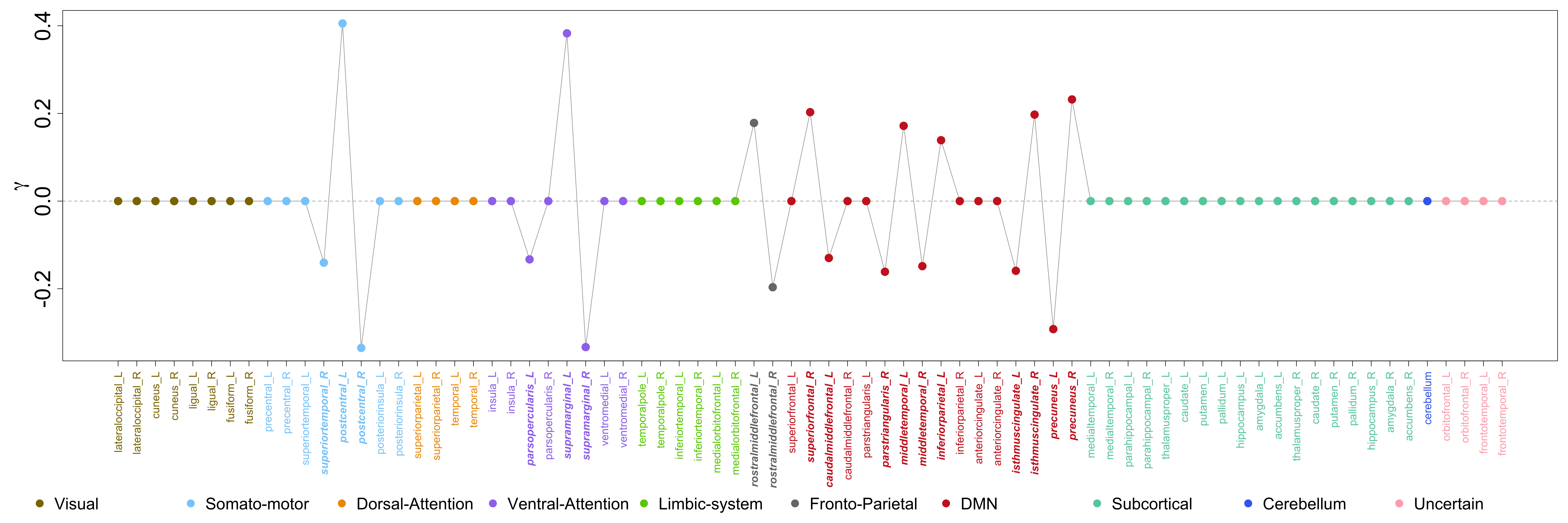}}

    \subfloat[\label{fig:hcp_gamma_C3}C3]{\includegraphics[width=0.81\textwidth]{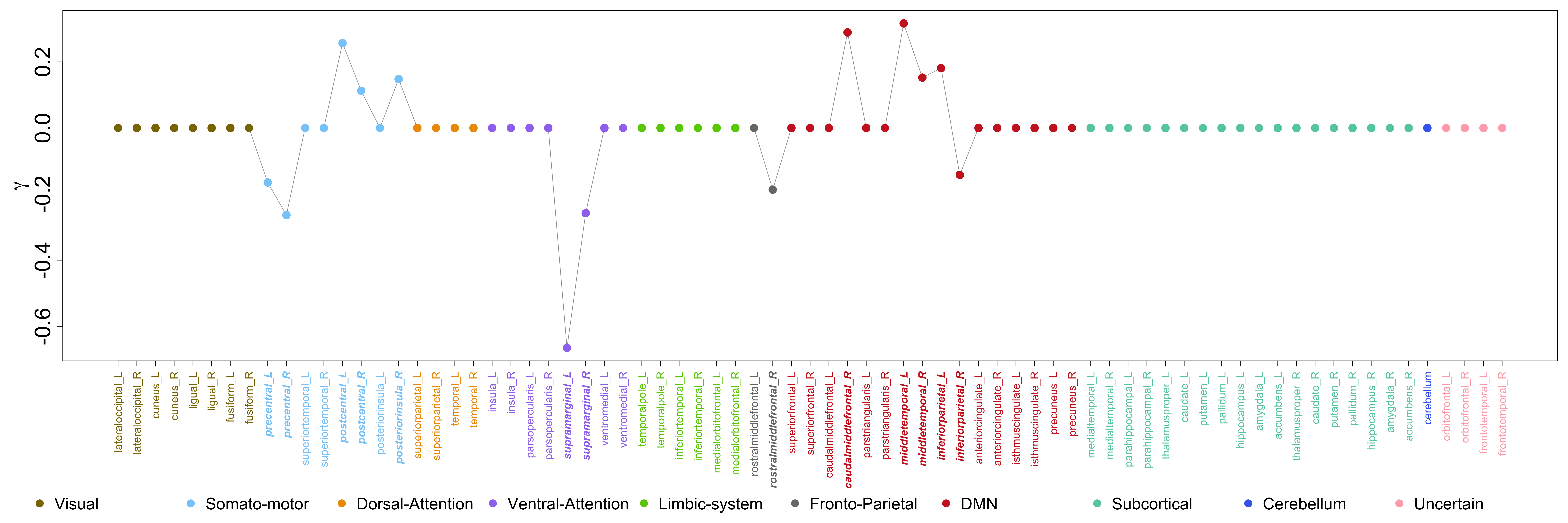}}
  \end{center}
  \caption{\label{fig:hcp_gamma}Sparsified loading profile ($\bgamma$) of the three identified components using the proposed approach in the HCP Aging study.}
\end{figure}
\begin{figure}
  \begin{center}
    \subfloat[\label{fig:hcp_beta_C1}C1]{\includegraphics[width=0.81\textwidth]{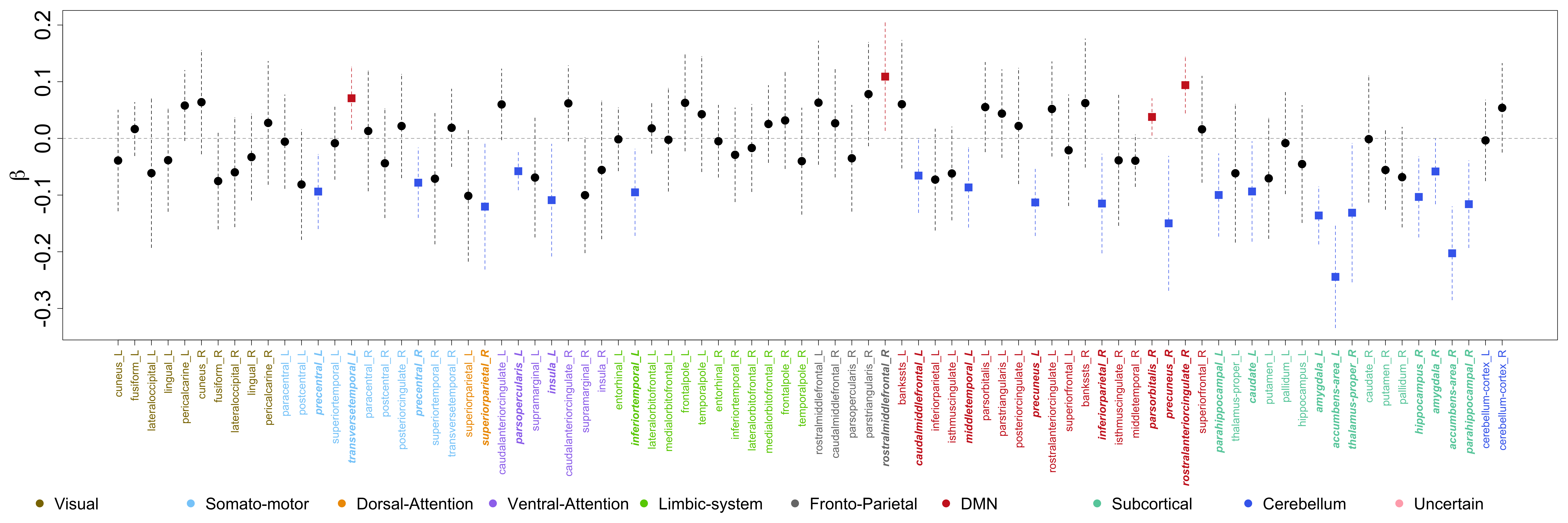}}

    \subfloat[\label{fig:hcp_beta_C2}C2]{\includegraphics[width=0.81\textwidth]{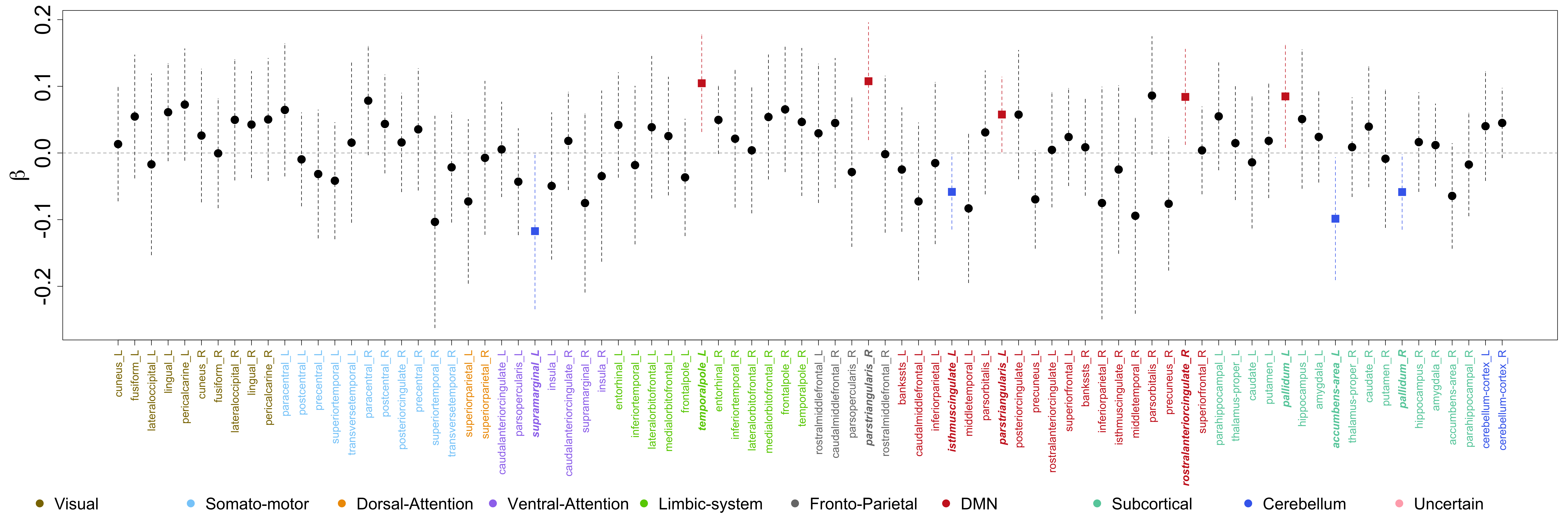}}

    \subfloat[\label{fig:hcp_beta_C3}C3]{\includegraphics[width=0.81\textwidth]{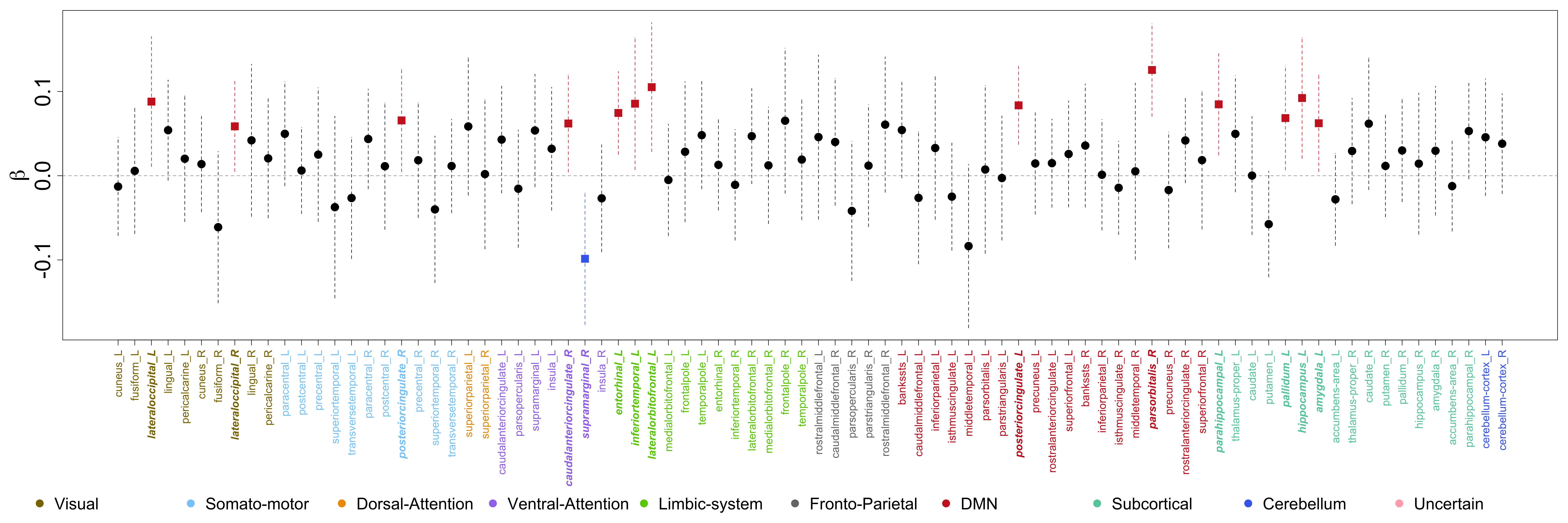}}
  \end{center}
  \caption{\label{fig:hcp_beta}Estimated model coefficient and $95\%$ confidence interval of the three identified components using the proposed approach in the HCP Aging study.}
\end{figure}

\begin{figure}
  \begin{center}
    \subfloat[\label{fig:hcp_river_gamma_C1}C1 rs-fMRI ($\bgamma$)]{\includegraphics[width=0.33\textwidth]{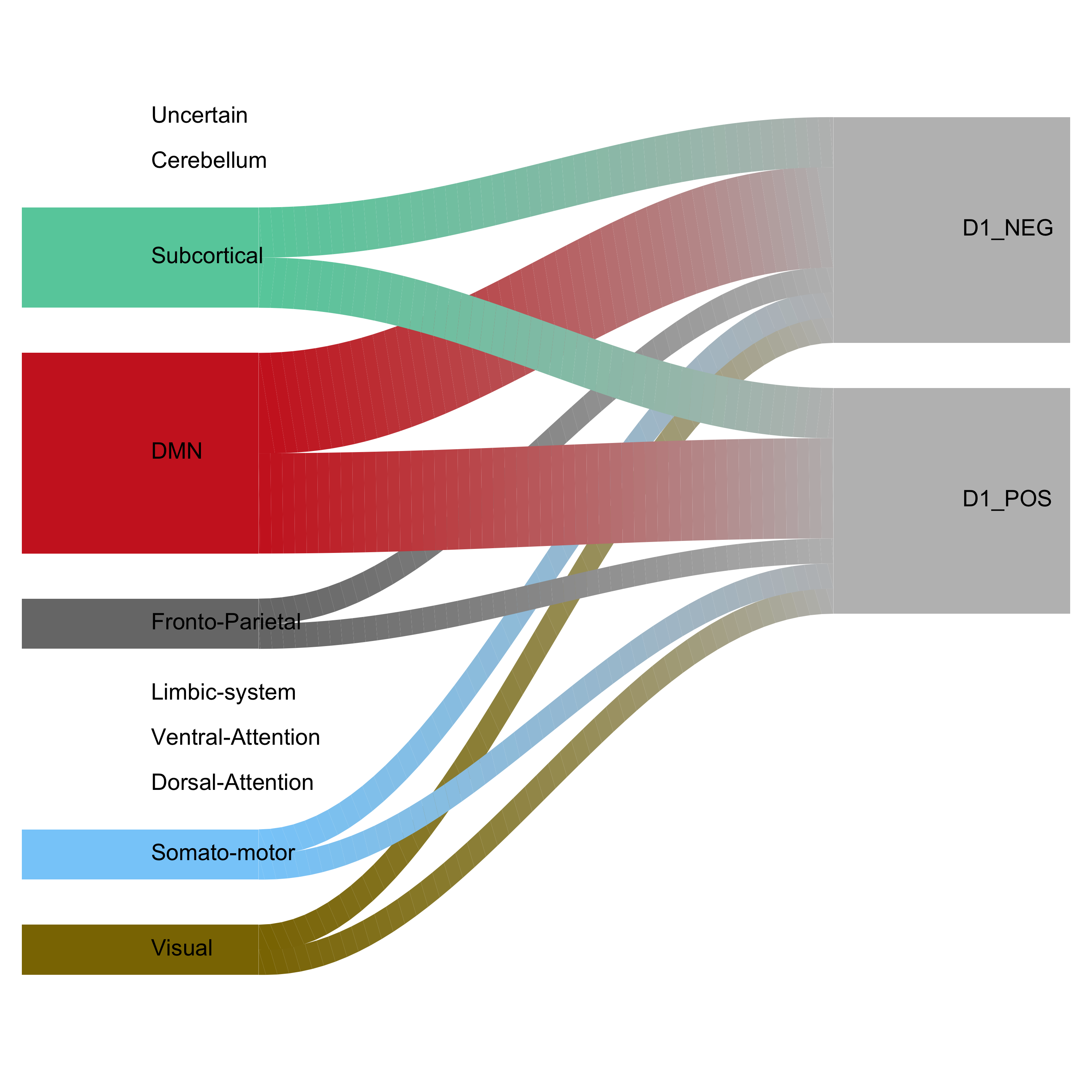}}
    \subfloat[\label{fig:hcp_river_gamma_C2}C2 rs-fMRI ($\bgamma$)]{\includegraphics[width=0.33\textwidth]{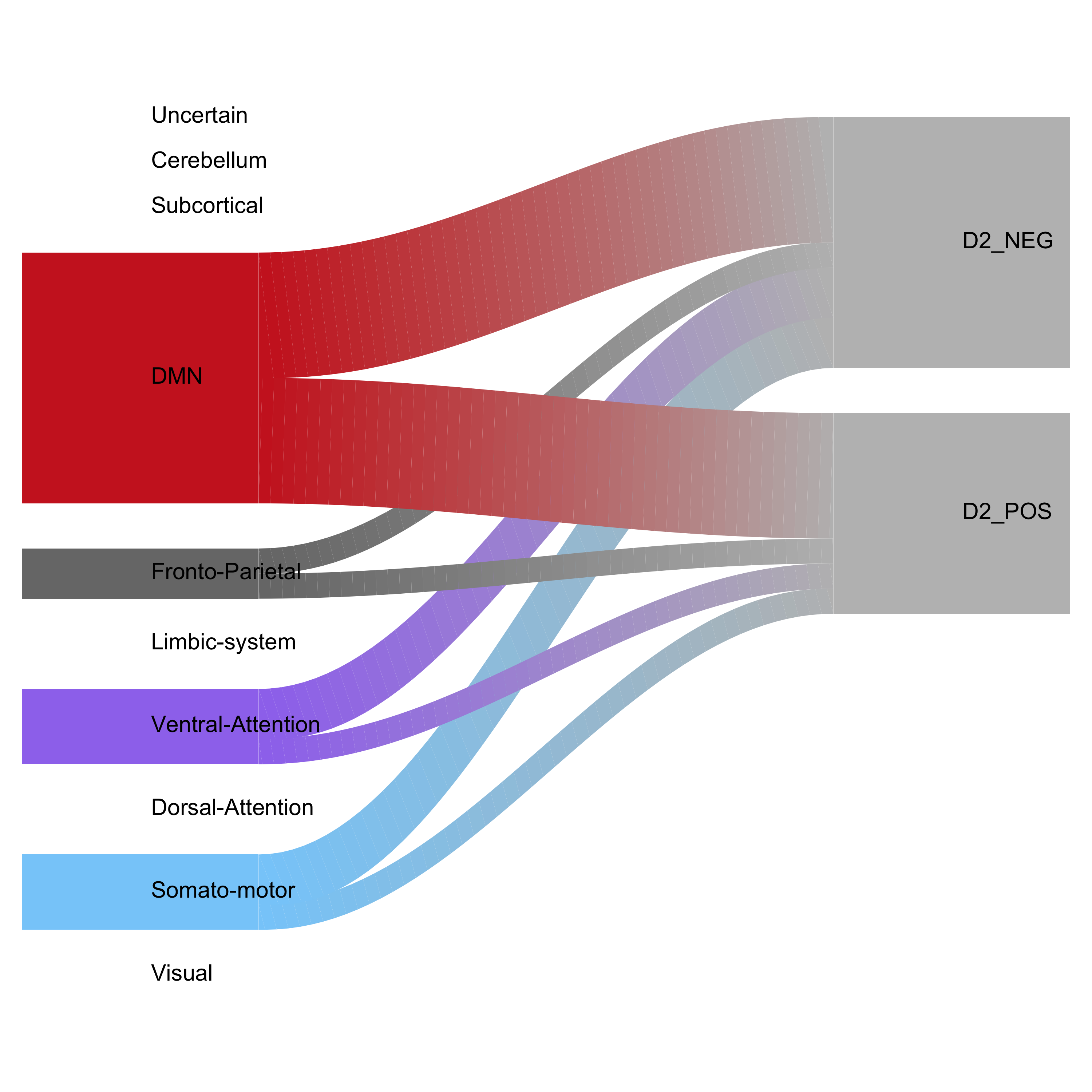}}
    \subfloat[\label{fig:hcp_river_gamma_C3}C3 rs-fMRI ($\bgamma$)]{\includegraphics[width=0.33\textwidth]{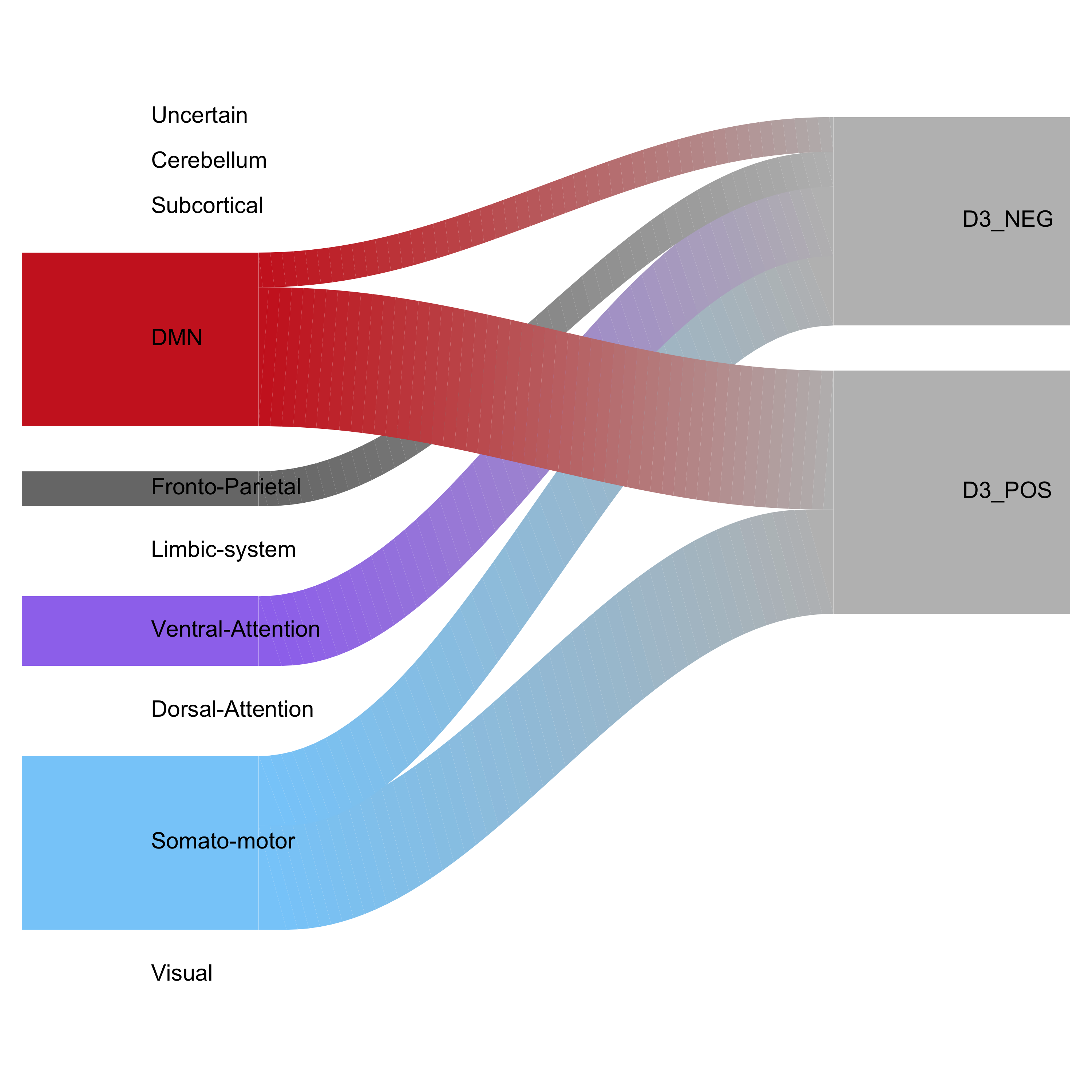}}

    \subfloat[\label{fig:hcp_river_beta_C1}C1 DTI ($\beta$)]{\includegraphics[width=0.33\textwidth]{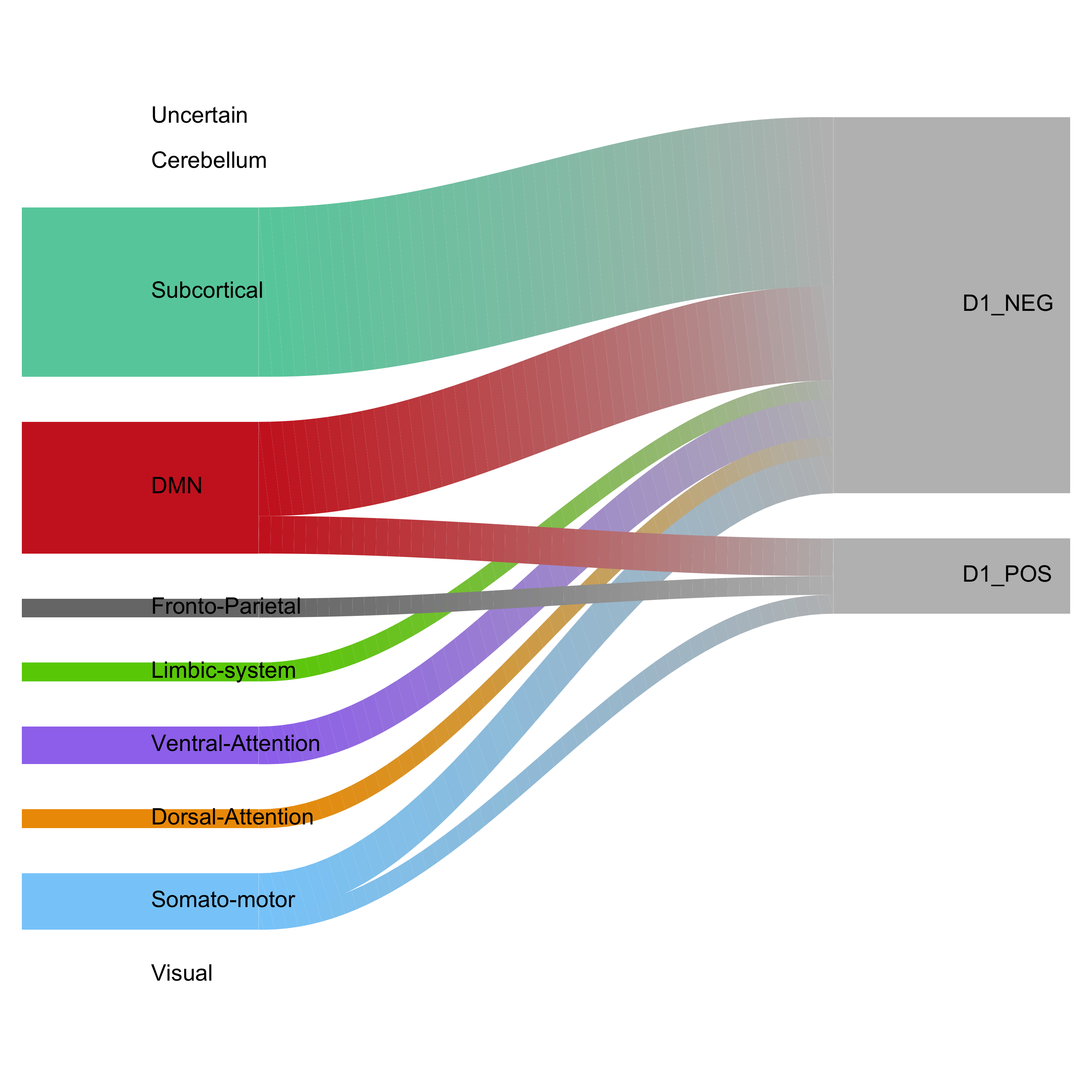}}
    \subfloat[\label{fig:hcp_river_beta_C2}C2 DTI ($\beta$)]{\includegraphics[width=0.33\textwidth]{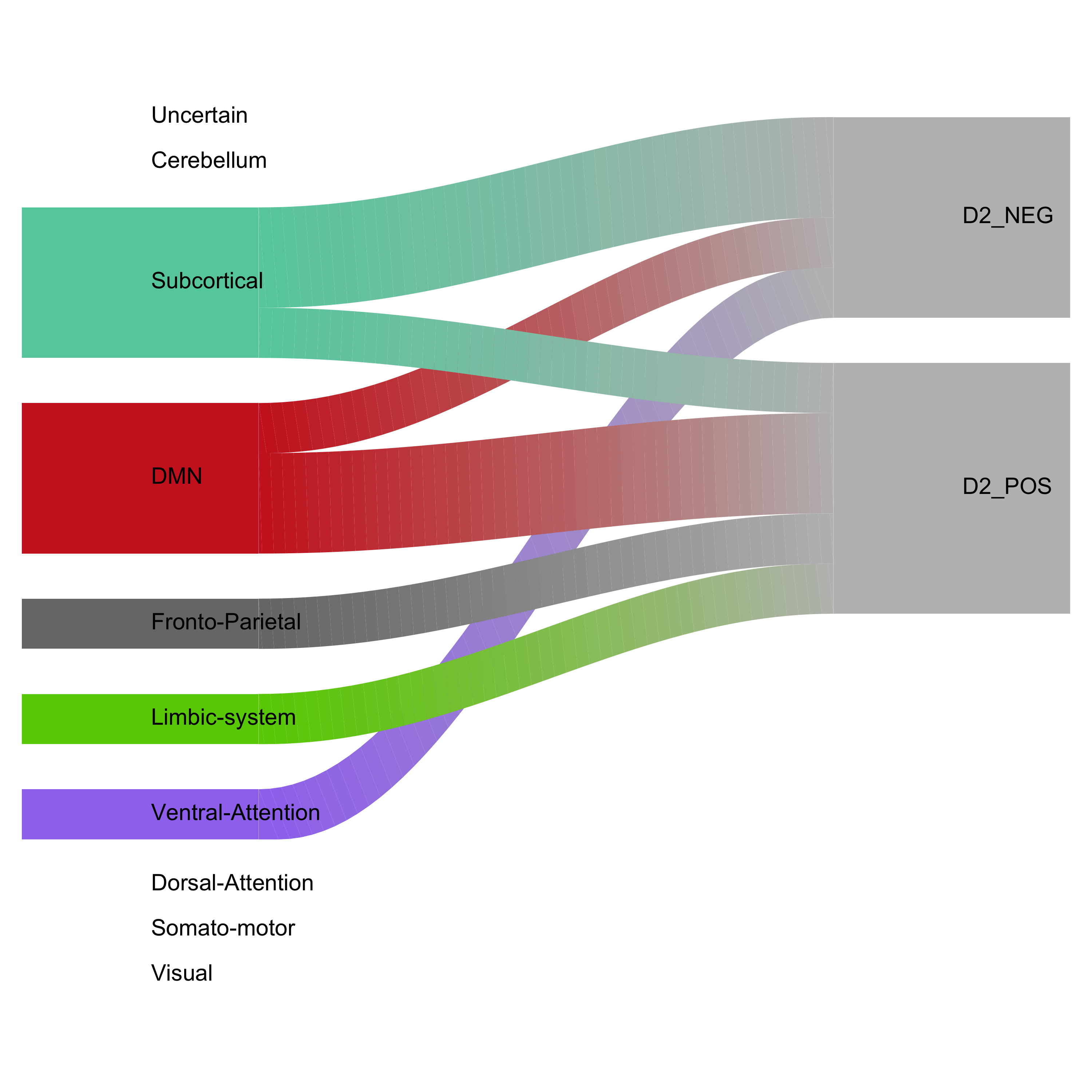}}
    \subfloat[\label{fig:hcp_river_beta_C3}C3 DTI ($\beta$)]{\includegraphics[width=0.33\textwidth]{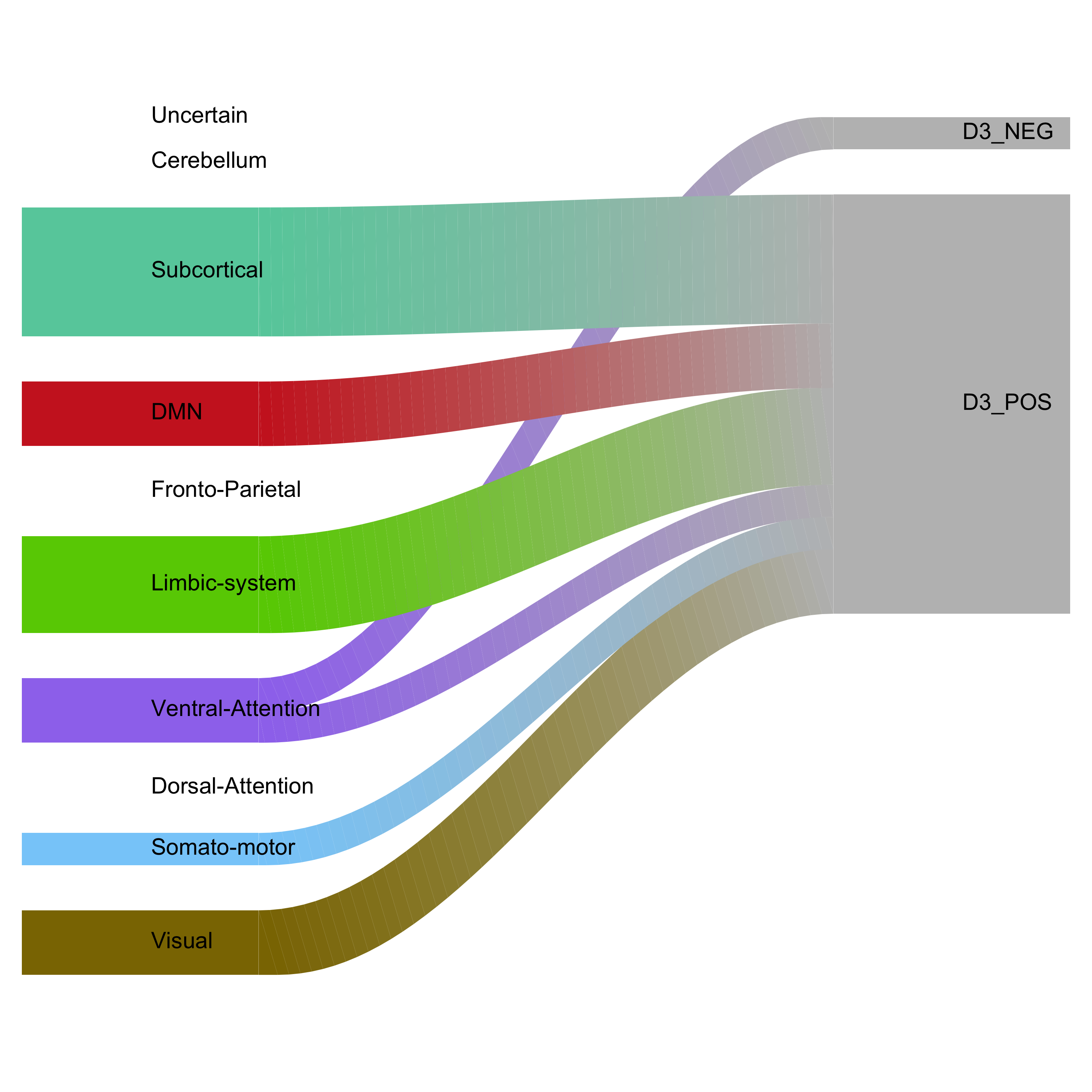}}
  \end{center}
  \caption{\label{fig:hcp_river}River plot of module configuration of the three identified components using the proposed approach in the HCP Aging study.}
\end{figure}
%========================================================%
%%%%%%%%%%%%%%%%%%%%%%%%%%%%%%%%%%%%%%%%%%%%%%%%%%%%%%%%%%

%%%%%%%%%%%%%%%%%%%%%%%%%%%%%%%%%%%%%%%%%%%%%%%%%%%%%%%%%%
%========================================================%
% Reference
%========================================================%
\clearpage

\bibliographystyle{apalike}
\bibliography{ref}
%========================================================%
%%%%%%%%%%%%%%%%%%%%%%%%%%%%%%%%%%%%%%%%%%%%%%%%%%%%%%%%%%

\end{document}